\documentclass[12pt]{iopart}
\pdfoutput=1 

\usepackage{color}
\usepackage{graphicx}
\usepackage{amssymb}
\expandafter\let\csname equation*\endcsname\relax
\expandafter\let\csname endequation*\endcsname\relax
\usepackage{amsmath}
\usepackage{amstext}
\usepackage{amsfonts}
\usepackage[caption=false]{subfig}
\usepackage[numbers,sort&compress]{natbib}
\usepackage{tikz}
\usepackage[bookmarks, bookmarksopen, bookmarksnumbered, colorlinks]{hyperref}
\usetikzlibrary{decorations.pathmorphing,calc}
\usetikzlibrary{decorations.pathmorphing}
\tikzset{font={\fontsize{22pt}{12}\selectfont}}

\newcommand{\half}{\kern.083333em}   
\newcommand{\quart}{\kern.0416675em}  
\newcommand{\nhalf}{\kern-.083333em}   
\newcommand{\nquart}{\kern-.0416675em}  

\newcommand\fscalar[1]{{}^\circ{\! #1}}
\newcommand\fvec[1]{{}^\dagger{\nhalf\nquart #1}} 
\newcommand\ftensor[1]{{}^\ddagger{\nquart #1}}

\newcommand\fperp[1]{\half{}^\perp{\nhalf #1}}
\newcommand\fpar[1]{\half{}^\parallel{\nhalf #1}}

\newcommand\curl{\mathop{\rm curl}\nolimits}
\newcommand\curlop[1]{\curl\nhalf #1} 
\newcommand\curlH{\curlop{H}}
\newcommand\curlE{\curlop{E}}
\newcommand\curlsigma{\curlop{\sigma}}

\renewcommand\div{\mathop{\rm div}\nolimits}

\newcommand\sgn{\mathop{\rm sgn}\nolimits}

\renewcommand{\d}{\text{d}} 

\newcommand\Rthree{R${}^3$~} 
\newcommand\Rfour{R${}^4$} 
\newcommand\Hthree{H${}^3$~} 
\newcommand\Tthree{T${}^3$~} 
\newcommand\Sthree{S${}^3$~} 

\newcommand{\ZZ}{\ensuremath{{\mathbb{Z}}}}

\newcommand{\threeR}{\mathcal{R}}

\newcommand{\Cscr}{\mathcal{C}}
\newcommand{\Sscr}{\mathcal{S}}
\newcommand{\Mscr}{\mathcal{M}}

\newcommand\dualC{{}^\star\nhalf C}

\newcommand{\eeff}{\ensuremath{{\rm eff}}}
\newcommand{\tot}{\ensuremath{{\rm tot}}}

\newcommand{\Lie}{{\ensuremath{\cal L}}}

\newcommand{\const}{\textrm{const}}
\newcommand{\UD}[2]{\ensuremath{^{#1}_{\phantom{#1} #2}}}

\newcommand{\beq}{\begin{equation}}
\newcommand{\eeq}{\end{equation}}
\newcommand{\bea}{\begin{eqnarray}}
\newcommand{\eea}{\end{eqnarray}}
\newcommand{\bit}{\begin{itemize}}
\newcommand{\eit}{\end{itemize}}
\newcommand{\bfi}{\begin{figure}}
\newcommand{\efi}{\end{figure}}
\newcommand{\bfic}{\begin{figure*}}
\newcommand{\efic}{\end{figure*}}
\newcommand{\bce}{\begin{center}}
\newcommand{\ece}{\end{center}}
\newcommand{\bt}{\begin{table}}
\newcommand{\et}{\end{table}}
\newcommand{\btb}{\begin{tabular}}
\newcommand{\etb}{\end{tabular}}
\newcommand{\todo}[1]{{\color{blue}~\textsf{[TODO: #1]}}}

\newcommand{\qed}{\nobreak \ifvmode \relax \else
      \ifdim\lastskip<1.5em \hskip-\lastskip
      \hskip1.5em plus0em minus0.5em \fi \nobreak
      \vrule height0.75em width0.5em depth0.25em\fi}

\begin{document}
\title{Black-Hole Lattices as Cosmological Models}

\author{Eloisa~Bentivegna$^1$, Timothy~Clifton$^2$, Jessie~Durk$^2$, Miko{\l}aj~Korzy\'{n}ski$^3$, and Kjell~Rosquist$^4$}
\address{
$^1$INFN, Sezione di Catania, Italy\\
$^2$School of Physics and Astronomy, Queen Mary University of London, UK\\
$^3$Center for Theoretical Physics, Polish Academy of Sciences, Warsaw, Poland\\
$^4$Department of Physics, Stockholm University, Sweden
}
\ead{eloisa.bentivegna@ct.infn.it, t.clifton@qmul.ac.uk, j.durk@qmul.ac.uk, korzynski@cft.edu.pl, kr@fysik.su.se}

\begin{abstract}
The search for solutions of Einstein's equations representing relativistic 
cosmological models with a discrete matter content has been remarkably fruitful in the last 
decade. 
In this review we discuss the progress made in the study of a specific 
subclass of discrete cosmologies, Black-Hole Lattice models. 
In particular, we illustrate the techniques used for the construction of
these spacetimes, and examine their resulting physical properties. This 
includes their large-scale dynamics, the dressing of mass due to the interaction
between individual black holes, along with features of direct observational
interest such as the distance-to-redshift relation.
This collection of results
provides a novel perspective on the physical effects of averaging in General Relativity,
as well as on the emergence of gravitational structures from solutions with isolated objects.
\end{abstract}

\setcounter{tocdepth}{2}
\tableofcontents

\section{Introduction}

Constructing a realistic model of the Universe is arguably one of the most daunting scientific tasks ever pursued, even when only standard stress-energy components are allowed. This is because the development of an effective cosmological theory, that is valid at the largest observable scales and that describes the (coarse-grained) consequences of deep hierarchies of gravitational systems, requires the adoption of many approximations and vast computational power~\cite{matarrese2011dark,aarseth2003gravitational,Anninos:2001yc}. The challenges posed in fully understanding these hierarchies, and their consequences for the large-scale properties of the Universe, are exponentially exacerbated by the fact that in General Relativity (GR) the gravitational interaction is governed by nonlinear field equations, with non-trivial gauge freedoms.

Though the metric of an isolated gravitational system was one of the first solutions of Einstein's equations to be discovered, finding the gravitational field corresponding to many-body systems is still something of an open problem in GR (outside of specific regimes~\cite{Chrusciel:2010my,Poisson:2011nh}). Unlike in Newton's gravitational theory, the superposition principle does not hold, and the nonlinear effects of combining several gravitating systems together have to be addressed on a case-by-case basis. Thanks to analytical and numerical techniques~\cite{Blanchet:2011wga,Damour:2012mv,Pretorius:2007nq}, great progress has recently been made on the two-body problem: solutions corresponding to compact-object binaries have been worked out in great detail for a variety of configurations. These allow for both the investigation of the fundamental nature of body-to-body interactions in GR, as well as for the determination of solutions to applied problems such as the modelling of gravitational-wave emission in the detection band of current and future interferometers~\cite{Buonanno:2014aza}.

In contrast, systems consisting of more than two bodies have only been explored relatively sparsely. The dynamics of three black holes were studied using Numerical Relativity (NR) in ~\cite{Diener:2003jc,Campanelli:2007ea,Campanelli:2005kr,Lousto:2007rj,Galaviz:2010mx}, whilst a general method to construct initial data for an arbitrary number of black holes was proposed in~\cite{Brandt:1997fk} and used for time evolution in~\cite{BruegmannFB:1997}. Of course, analytical approaches based on post-Newtonian expansions also exist, and have been used to capture the dynamics of multiple bodies in GR up to 3.5PN order~\cite{Futamase:2007zz}. Both direct simulations and analytical frameworks hint at a rich phenomenology, which parallels and extends that which is observed in the non-relativistic case (for instance, in the existence of choreographic  order~\cite{Imai:2007gn}).

Naturally, matter in the real Universe is much more complex than these few-body solutions. In addition, studies of spacetimes filled with a finite number of bodies are typically limited to spaces which are asymptotically-flat, and are therefore unfit to describe the real large-scale Universe. These deficits have been bridged in the past decade by the construction of models that contain arrangements of structures built from the basic one-body building block of GR (the Schwarzschild solution), but which exhibit homogeneity and isotropy on large scales. These {\it Black-Hole Lattices} (BHLs), as they have been dubbed, have made use of both analytic approaches (in the form of post-Newtonian and perturbative studies) and direct numerical integration. They have shed light on the allowed configurations for regular arrangements of black holes, as well as their dynamics, their continuum and weak-field limits, their optical properties, and their gravitoradiative content. They are the focus of this review.

The study of discrete, strongly nonlinear cosmologies offers an interesting perspective on the role of coarse graining in General Relativity. This problem involves, first and foremost, the question whether averaging an inhomogeneous cosmology leads to a sensible model in the FLRW class. It is critical to remark that the answer to this question may well depend on the particular observable under consideration (i.e., an inhomogeneous model may exhibit the same large-scale expansion as some FLRW model, but the average matter or curvature evolution of another). Any effective theory of cosmological inhomogeneities must therefore entail a number of different maps to the FLRW class, each of which should be studied separately. Below we will review how this has so far been achieved within the class of BHL models for phenomena such as the evolution of the scale factor (Section~\ref{sec:evol}), the dressing of the mass parameters (Section~\ref{sec:effect}) and the optical properties of the spacetime (Section~\ref{sec:opt}). It is important to remark that the question of which averaged cosmology best fits a given inhomogeneous spacetime does not necessarily have a unique answer, and that different phenomena can (and should, in general) be mapped to different FLRW models.

In this article, we review recent progress in the construction and analysis of BHLs. In Section~\ref{sec:field_eqns} we outline Einstein's equations and their 3+1 decomposition (on which many of the relevant studies are based). In Section~\ref{sec:arr} we articulate the construction of initial data in different scenarios, highlighting where exact prescriptions can be made, as well as the role of symmetries. In Section~\ref{sec:evol} we outline the current analytical and numerical approaches for the time development of BHL initial data. Section~\ref{sec:cont} contains a discussion of the various techniques, analytical and approximate, used to study the continuum and weak-field limits of these cosmological models. In Section~\ref{sec:effect} we explain why relativistic interactions can mean that a BHL can exhibit different effective properties from its individual building blocks, and illustrate this with a controlled study of the effects of structure formation on the large-scale properties of the spacetime. Finally, in Section~\ref{sec:opt} we describe the study of light propagation in these models, which is a subject of critical importance from the observational standpoint. Throughout the paper, we use geometric units where $8 \pi G = c = 1$. We also adopt the convention of reserving Latin letters from the first half of the alphabet for abstract indices, while using those from the second half to denote spatial tensor components. 

\section{Field equations}
\label{sec:field_eqns}
Recent studies of universes with discrete matter have relied on a 3+1 formulation of the Einstein's field equations~\cite{alcubierre2008introduction}. The first step in applying this formulation involves finding a solution of the constraint equations on a given spatial hypersurface $\Sscr_0$. This solution can then be used as initial data for a numerical and/or analytical study of the evolution of the system. Applying the evolution equations will then give a spacetime which can be considered as foliated by a sequence of hypersurfaces $\Sscr_t$ labeled by a time function $t$.
If $g_{ab}$ is the spacetime metric, then the hypersurfaces $\Sscr_t$ have an induced spatial metric  
\begin{equation}
   h_{ab} = g_{ab} + u_a u_b
\end{equation}
where $u^a$ is the unit normal of $\Sscr_t$. As $u^a$ is a unit timelike vector field it can be interpreted as the 4-velocity of observers at rest with respect to $\Sscr_t$ (a fact that is especially relevant for the cosmological applications considered in this review).

Covariant derivatives can be decomposed in the 3+1 approach into temporal and spatial parts relative to $u^a$ by defining
\begin{equation}
 \begin{split}
  \dot S^{a\ldots b}{}_{c\ldots d} 
   &= u^e\nabla_e S^{a\ldots b}{}_{c\ldots d} \\[3pt]
  D_e S^{a\ldots b}{}_{c\ldots d}
   &= h^a{}_i\ldots h^b{}_j \half h_c{}^m\ldots h_d{}^n
        \half h_e{}^f \half\nabla_f S^{i\ldots j}{}_{m\ldots n} \, .
 \end{split}
\end{equation}
Besides the intrinsic geometry $h_{ab}$ of $\Sscr_t$, the hypersurfaces are also characterised by how they are immersed in the spacetime. This can be quantified by the extrinsic curvature, defined by
\begin{equation}
   K_{ab} = -h_a{}^c h_b{}^d \half\nabla_{(c} u_{d)} \, .
\end{equation}
Here we consider Einstein's equations $G_{ab}= T_{ab}$ with the following stress-energy tensors:
\begin{itemize}
\item 
\emph{Vacuum:} $T_{ab}=0$
\item
\emph{Cosmological constant:} $T_{ab}= -\Lambda g_{ab}$
\item
\emph{Electrovacuum:} $T_{ab}= F_{ac} F_b{}^c - \frac14 g_{ab} F_{cd} F^{cd}$
\item
\emph{Free scalar field:} $T_{ab} = \nabla_a\varphi  \nabla_b\varphi
                   - \frac12 g_{ab} \nabla_c\varphi  \nabla^c\varphi$
\end{itemize}
These stress-energy tensors can be decomposed into quantities measured by an observer following the integral curves of $u^a$, and who is hence at rest with respect to $\Sscr_t$, as follows:
\begin{equation}
 \begin{split}
     \rho &= u^a u^b T_{ab} \quad \quad \, \text{energy density} \\
        p &= \tfrac13 h^{ab} T_{ab} \quad \quad \, \text{pressure}  \\
      q_a &= -h_a{}^b u^c \half T_{bc} \quad
             \text{energy flux (or equivalently momentum density)} \\
    \pi_{ab} &= T_{\langle ab \rangle} \quad \quad \quad \; \;
             \text{anisotropic pressure and tangential stresses}
 \end{split}
\end{equation}
Here the angle brackets indicate the spatial tracefree part of any symmetric  pair of indices, such that
\begin{equation}
   S_{\langle ab \rangle}
    = \bigl(h_{(a}{}^c h_{b)}{}^d - \tfrac13 h_{ab} \, h^{cd}\bigr) S_{cd} \, .
\end{equation}
Einstein's equations can also be decomposed into their components parallel and orthogonal
to $u^a$. One thus obtains the constraint equations, which have the form
\begin{align}\label{scalar_constraint}
   \threeR + K^2 - K_{ab}K^{ab} &= 2\rho  \\ \label{mom_constraint}
            D_b K^b{}_a - D_a K &= 2q_a
\end{align}
where $\threeR$ is the intrinsic Ricci scalar of $\Sscr_0$ and $K= h^{ab} K_{ab}$ is its mean curvature. The right-hand sides of these equations represent any non-gravitational fields that are present, including also the cosmological constant (if present).

The flow of $u^a$ can be taken to be geodesic, and can be used to define a synchronous time variable. However, while such a choice of time is simple from the analytic point of view, it can eventually become problematic, as the focusing of the $u^a$ congruence typically means that caustics develop in finite time. To avoid focusing of the timelike congruence, one can use an alternative 3+1 slicing by defining a timelike vector $w^a$ 
\begin{equation}
   w^a = \alpha u^a + \beta^a
\end{equation}
such that $w^a t_{,a}=1$ and where $\alpha = 1/ (u^a t_{,a})$.
We can then generate the spacetime by evolving the spatial metric and the  extrinsic curvature using the remaining Einstein's equations, which take the form
\begin{equation}\label{eq:adm}
 \begin{split}
   \dot h_{ab} + 2\alpha K_{ab} - 2D_{(a}\beta_{b)} &= 0 \\
   \dot K_{ab} + \alpha (2K_{ac} K^c{}_b - K K_{ab} - \threeR_{ab})
               + \Lie_\beta K_{ab} - D_a D_b \half\alpha
     &= \tfrac12 (\rho - p) h_{ab} + \pi_{ab}
 \end{split}
\end{equation}
where $\threeR_{ab}$ is the intrinsic Ricci tensor of $\Sscr_t$. While this system takes on the form of an initial-boundary-value problem, and therefore appears ready to describe the evolution of given initial-data sets in a chosen time direction, it has long been known in practice that the numerical integration of (\ref{eq:adm}) is not well-behaved~\cite{alcubierre2008introduction}. 

A widely-adopted reformulation of this system that cures the numerical instabilities is referred to as the Baumgarte-Shapiro-Shibata-Nakamura (BSSN)~\cite{Nakamura:1987zz, Shibata:1995we, Baumgarte:1998te}
system, and takes the form
\beq
\begin{split}
(\partial_t - \beta^a \partial_a) W &= - \frac{1}{3} \alpha K + \frac{1}{3} \partial_a \beta^a  \\
(\partial_t - \beta^a \partial_a) K &= - D_a D^a \alpha + \alpha (\bar A_{ab} \bar A^{ab} + \frac{1}{3} K^2)  \\
(\partial_t - \beta^a \partial_a) \bar h_{bc} &= -2 \alpha \bar A_{bc} + 2 \bar h_{a(b}\partial_{c)} \beta^a - \frac{2}{3} \bar h_{bc} \partial_a \beta^a  \\
(\partial_t - \beta^a \partial_a) \bar A_{bc} &= W^2 (-D_{\langle b} D_{c \rangle} \alpha + a \threeR_{\langle bc \rangle})  \\
  & + \alpha (K \bar A_{bc} - 2 \bar A_{ba} \bar A^a_c) \label{eq:bssn} 
  + 2 \bar A_{a(b} \partial_{c)} \beta^a - \frac{2}{3} A_{bc} \partial_a \beta^a  \\
(\partial_t - \beta^a \partial_a) \bar \Gamma^b &= \bar h^{ac} \beta^b \partial_a \beta_c +
    \frac{1}{3} \bar h^{ab} \partial_b \partial_c \beta^c - \bar \Gamma^a \partial_a \beta^b  \\
  &+ \frac{2}{3} \bar \Gamma^b \partial_a \beta^a
      - 2 \bar A^{ab} \partial_a \alpha  
      + 2 \alpha (\bar \Gamma^b_{ac} \bar A^{ac} - 3 \bar A^{bc} \partial_c \ln W - \frac{2}{3} \bar h^{bc} \partial_c K)  
\end{split}
\eeq
where $W=\det h^{-1/6}$ and $\bar \Gamma^a=-\partial_b \, \bar h^{ab}$ are auxiliary variables.
All numerical simulations described in this review are based on the integration of a finite-difference
discretisation of the system (\ref{eq:bssn}), along with the following gauge prescription~\cite{Campanelli:2005dd,Baker:2006yw}:
\bea
(\partial_t - \beta^a \partial_a) \alpha &=& - 2 \alpha K \\
(\partial_t - \beta^a \partial_a) \beta^b &=& \frac{3}{4} B^b \\
(\partial_t - \beta^a \partial_a) B^b &=& - (\partial_t - \beta^a \partial_a) \bar \Gamma^b - \eta B^b \, .
\eea
This coordinate choice is known as the \emph{moving-puncture} gauge.

The reader may note that in the literature of theoretical cosmology it is common to use a terminology closely related to that one described above, but which more directly refers to the physical properties of the velocity field $u^a$. In this alternative approach the basic kinematical variables are taken to be the (volume) expansion $\Theta = \nabla_a u^a = -K$ and the shear $\sigma_{ab}= -A_{ab}$, where $A_{ab} = K_{\langle ab\rangle}$ is the spatial tracefree part of the extrinsic curvature. The dynamical field variables are given by the Weyl curvature, which can be decomposed into gravitoelectric and gravitomagnetic parts as
\begin{equation}
   E_{ab} = u^a u^b C_{acbd} \ ,\qquad H_{ab} = u^a u^b \,\dualC_{acbd}
\end{equation}
where $\dualC_{acbd} = \frac12 \epsilon_{ab}{}^{ef} C_{efcd}$. This leads to field equations which involve quantities with manifest physical interpretation \cite{Maartens:1997}.
To display those equations in a concise form it is convenient to introduce shorthand notations for divergence and curl operations on symmetric second-rank spatial tensors by
\begin{equation}
   (\div\!S)_a = D^b S_{ab} \ ,\qquad (\curl\! S)_{ab} = \epsilon_{cd(a} D^c S_{b)}{}^d
\end{equation}
and the bracket operations
\begin{equation}
   \langle S, P\rangle = S_{c\langle a} P_{b\rangle}{}^c \ ,\qquad
   [S,P]_a = \epsilon_{abc} S^b{}_d P^{cd} \, .
\end{equation}
The evolution equations then become
\begin{equation}
\label{m_ev}
 \begin{split}
        \dot\Theta &= -\tfrac13 \Theta^2 - \sigma_{ab}\half\sigma^{ab}\\[3pt]
   \dot\sigma_{ab} &= -\tfrac23 \Theta -\langle\sigma,\sigma\rangle_{ab}
                      -E_{ab} \\[3pt]
       \dot E_{ab} &= -\Theta E_{ab} + 3\langle\sigma,E\rangle_{ab}
                      +\curlH_{ab} \\[3pt]
       \dot H_{ab} &= -\Theta H_{ab} + 3\langle\sigma,H\rangle_{ab}
                      -\curlE_{ab} \, ,
 \end{split}
\end{equation}
which are subject to the constraint equations
\begin{equation}
\label{m_constr}
 \begin{split}
             (\div\!\sigma)_a &= \tfrac23 D_a\Theta \\[3pt]
                  (\div\!E)_a &=  [\sigma,H]_a \\[3pt]
                  (\div\!H)_a &= -[\sigma,E]_a \\[3pt]
   (\curl\!\sigma\nhalf)_{ab} &= H_{ab} \, .
 \end{split}
\end{equation}
This formulation of Einstein's equations is useful for extracting analytic information about the evolution of cosmological spacetimes, and can be used to help understand the evolution of BHLs.

\subsection{Averaged spacetimes and backreaction}
\label{sec:ave}
Evaluating the effect of coarse-graining spacetime in cosmology is one of the original motivations for studying black-hole lattices. The underlying problem here is that Einstein's equations describe the local Ricci curvature of spacetime, and not the coarse-grained average. If we have a matter distribution that is close to being homogeneous and isotropic on large scales, but is inhomogeneous on smaller scales, it is therefore {\it not} necessarily the case that the large-scale evolution of the cosmological average is given by applying Einstein's equations to FLRW geometries with simple matter contents.

The natural thing to do in this situation is to introduce an idealised, coarse-grained (or averaged) metric tensor $g^{(0)}$, which does not contain the contribution from the smallest inhomogeneities, but 
correctly encodes the large-scale geometry of the full solution of Einstein's equations. We can imagine that this metric satisfies the effective, or averaged Einstein's equations,
\bea
G_{ab}\left[g^{(0)}\right] = T^{(0)}_{ab}\, ,
\eea
where $ T^{(0)}_{ab}$ is defined as the {\it effective} stress-energy tensor. Note, however, that this stress-energy tensor will in general be different than the simple average of the local one, i.e. $T_{ab} \neq \left\langle T_{ab}\right\rangle$ (where angular brackets here denote a coarse-grained or averaged value). The difference between the two is often referred in the literature as the 
\emph{backreaction} of inhomogeneities on the large-scale expansion of space \cite{Buchert:2007ik}. 

The backreaction term in Einstein's field equations can be made explicit by defining $B_{ab} = T^{(0)}_{ab} - \left\langle T_{ab}\right\rangle$, such that
\bea
G_{ab}\left[g^{(0)}\right] = \left\langle T_{ab}\right\rangle + B_{ab} \, .
\eea
The \emph{backreaction tensor}, $B_{ab}$, measures the extent to which the large-scale degrees of freedom of the system fail to obey Einstein's equations with the ``naively'' averaged stress-energy tensor, and can be shown to appear due to the nonlinear structure of the equations. The issue of backreaction and its relevance in cosmology has sparked a long debate between researchers, with some believing the backreaction term to be negligible in the cosmological setting \cite{Ishibashi:2005sj, Green:2010qy, Green:2014aga, Green:2015bma} while others maintain that we lack sufficiently sound mathematical and physical arguments to be able to state such conclusions (see e.g. \cite{Buchert:2015iva}). For status reports on this topic the reader is referred to \cite{Buchert:2007ik, Bolejko:2016qku} for detailed reviews.

The reader may note that the definition of backreaction given above is inherently ambiguous: it requires us to specify (i) exactly how we define the idealised metric $g^{(0)}$ given the physical
one $g$ and (ii) how the stress-energy tensor should be averaged. Many methods have been proposed for these problems \cite{Korzynski:2009db,  Carfora:2008fr, Zalaletdinov:2008ts}, but the most commonly used formalism was introduced by Buchert \cite{Buchert:1999er, Buchert:2001sa}. In his approach the volume of a spatial domain in comoving coordinates is used to define the effective scale factor $a(t)$, which is then used to determine the effective energy density and backreaction via the effective Friedman equations (i.e. the coarse-grained Einstein's equations for a homogeneous and isotropic gravitational field). For our present purposes, it may be noted that if we perform a 3+1 decomposition on Einstein's equations then we may distinguish two types of backreaction effects. The first type is connected to the constraint equations and the $T\UD{0}{\mu}$ component of Einstein's equations. This does not involve dynamics, and it is therefore already possible to study it at the level of initial data (without needing to perform any numerical or approximate evolution of the system). The second type is a dynamical one, and is connected with the evolution, as specified by the ADM or BSSN equations above.

The BHLs reviewed in this report represent solutions in which the matter distribution is as clumped as possible on small scales, while being close to perfectly homogeneous and isotropic on large scales. They therefore serve as excellent testbeds for studying the backreaction problem: Averages or macroscopic measures of scale or expansion can be extracted with relative ease, and ideas involving the links between small and large-scale gravitational fields can be explored in detail.

\subsection{Vacuum initial data}
\label{sec:vacuum}
Finding suitable initial configurations for a BHL entails the solution of the constraint 
system (\ref{scalar_constraint})-(\ref{mom_constraint}). In black-hole spacetimes, it is often
more convenient to resort to a conformal decomposition,
which in vacuum takes the form
\beq
\begin{split}
\label{eq:CTTconstraints}
 \bar \Delta \psi - \frac{\bar \threeR}{8}\,\psi - \frac{K^2}{12}\,\psi^5 + \frac{1}{8} {\bar A}_{ij} {\bar A}^{ij} \psi^{-7} &= 0 \\
 \bar D_i \bar A^{ij} - \frac{2}{3} \psi^6 \bar \gamma^{ij} \bar D_i K &= 0
\end{split}
\eeq
where $\bar \Delta$ is the Laplacian operator of the conformal metric
$\bar h_{ij}$, defined by $h_{ij} = \psi^4 \bar h_{ij}$, $\bar \threeR$ is its scalar curvature,
and $\bar A_{ij}$ is related to $A_{ij}$ by
$\bar A_{ij}=\psi^2 A_{ij}$. In practice, this system is solved by choosing
free data (e.g.~for $\bar h_{ij}$, $K$, and some parts of $\bar A_{ij}$), and
using the four equations above to solve for the remaining variables.

An important detail to keep in mind is that, in periodic spaces, free data cannot be
chosen in a completely arbitrary way. Taking the integral of the first constraint
equation over the fundamental cell, it becomes apparent that if black holes are present,
then $\bar \threeR$, $K$ and $\bar A_{ij}$ cannot all vanish (for details of this argument
in the case of BHLs, see e.g.~\cite{Bentivegna:2013ata}, or~\cite{Kleban:2016sqm} for a 
more general discussion). This additional constraint implies, for instance, the 
non-existence of periodic, time-symmetric, conformally-flat initial data. 

This is also apparent if one tries to find periodic solutions to the constraint
equations.
In the case of backgrounds which have a compact universal covering space, this can be done without obstructions. However, the situation is different when the universal covering space is not compact. For example, if the background is flat space, then the constraint to be solved is the Laplace equation. A simple periodic distribution could be formally arranged by choosing a cubic lattice and putting one black hole at the centre of each cell. A corresponding solution of the Laplace equation would then be periodic in all three directions. However, each black hole represents a simple pole solution of the Laplace equation, and, as is well-known, there exists no periodic solution with simple poles having positive masses. In this context it is of interest to note that for a universe with a nonzero ordinary matter density to admit time symmetric initial data, it must have positive curvature as follows from \eqref{scalar_constraint}. 

Keeping this 
condition in mind, in the following subsections we describe two specific examples 
of initial-data constructions. 
The most straightforward ansatz is to use time-symmetric initial data corresponding to a configuration of Schwarzschild sources 
\cite{Brill:1963yv}. As we will see in Section~\ref{sec:momentarily}, the time-symmetric data is by definition characterised by the conditions $\Theta|_0=0$ and $\sigma|_0=0$, where the zero subscript refers to an  initial data surface given by $t=t_0$.  Based on the considerations above, this can only be done for a universe which has a turning point (the moment of time symmetry) where expansion is replaced by contraction. As we will see, this excludes an \Rthree\ background and also hyperbolic space \Hthree\ as well as topologies which have \Rthree\ or \Hthree\ as universal covering space, like the 3-torus \Tthree\ for example. 
It is, of course, also possible to drop the time-symmetry requirement and search for
solutions with, say, an \Rthree background. This, however, requires a numerical approach,
as described in Section~\ref{sec:expanding}.
 
\subsubsection{Momentarily static initial data in vacuum.}
\label{sec:momentarily}
The original starting point for cosmological lattice constructions \cite{Clifton:2012qh,Bentivegna:2012ei} is the same as that used for black hole collisions with two or more black holes initially at rest with respect to each other. The constraint equations for such an initial data configuration reduce to a single equation coming from the Hamiltonian (scalar) constraint \eqref{scalar_constraint}, which in vacuum becomes
\begin{equation}\label{ham_constraint}
   \threeR = -\textstyle\frac23 \Theta^2 + \sigma_{ab} \half\sigma^{ab}\end{equation}
where $\threeR$ is the scalar curvature of the 3-metric in the initial data, and the remaining terms contain the expansion and shear. Lattice models, that are built using black holes that satisfy this equation, are by definition vacuum universes. If we impose the conditions for instantaneous time symmetry,  
$\Theta=0$ and $\sigma_{ab} =0$, then the only remaining constraint is $\threeR =0$. Any initial data that satisfies this equation is sufficient to provide a unique evolution.

In the cosmological context, it is convenient to consider the isotropic form of the Schwarzschild solution, for which the metric takes the form
\begin{equation}\label{Schw}
   g_\text{S} = - \left( \frac{\psi_-}{\psi_+} \right)^2 \d t^2
            + \psi_+^4 \bigl(\d\tilde r^2 + \tilde r^2 \d\Omega^2 \bigr)
\end{equation}
where
\begin{equation}\label{psipm}
   \psi_\pm = 1 \pm \frac{m}{2\tilde r}
\end{equation}
and 
\begin{equation}     \d\Omega^2 = \d\theta^2 + \sin^2\!\theta \half\d\phi^2
\end{equation}
is the round 2-sphere metric. Each hypersurface of constant time $t$ in this geometry can then be taken to correspond to time-symmetric initial data. It is also apparent from Eq.\eqref{Schw} that any such initial data will be conformally flat, as this is a manifest property of the spatial part of this metric.

More generally, one may consider an auxiliary conformally rescaled metric for the initial data, given by
\begin{equation}
   \bar h_{ab} = \psi^{-4} h_{ab}
\end{equation}
where $h_{ab}$ is the physical 3-metric. In the following we will refer to $\bar h_{ab}$ as the background metric (or background geometry when referring to a corresponding global structure). 
The constraint \eqref{ham_constraint} now becomes
\begin{equation}\label{rescaled_constraint}
   \threeR = \psi^{-4}(\bar\threeR - 8\psi^{-1}\bar\Delta\psi) = 0
\end{equation}
where $\bar\threeR$ and $\bar\Delta$ are the scalar curvature and Laplace operator  corresponding to the auxiliary metric $\bar h_{ab}$.
For the Schwarzschild solution in the form \eqref{Schw}, $\psi=\psi_+$ and $\bar h_{ab}$ is flat implying $\bar\threeR=0$. The constraint \eqref{rescaled_constraint} then reduces to the flat space Laplace equation
\begin{equation}
   \bar\Delta\psi = 0
\end{equation}
for which \eqref{psipm} is a well-known solution. For a general conformally flat background, the constraint \eqref{rescaled_constraint} can be expressed in the form
\begin{equation}\label{genH}
   \bar\Delta\psi = \tfrac18 \bar\threeR \psi \, .
\end{equation}
If the background curvature $\threeR$ is constant, then this is the Helmholtz equation. However, for any background geometry, the generalised Helmholtz equation \eqref{genH} is still linear and homogeneous and therefore solutions can be superposed. This makes it possible to construct exact initial data consisting of a superposition of multiple Schwarzschild black holes each of which has an arbitrary mass and is at an arbitrary position in the background geometry. 
As described in Section~\ref{ssec_extra}, the above construction can be generalised to include a cosmological constant, charged black holes and matter in the form of a scalar field.

For a round 3-sphere as the background metric the solution for $\psi$ in \eqref{genH} can be expressed in a closed form as
\bea
\psi(x) = \sum_{i=1}^N \frac{\alpha_i}{\sin \left( \frac{\lambda(x,x_i)}{2} \right) } \, , \label{eq:psiclosedform}
\eea
where $x_i$ are $N$ distinct points in \Sthree, the $\alpha_i$ are positive superposition parameters, and $\lambda(\cdot,\cdot)$ is the standard distance on the unit round 3-sphere. This conformal factor can be seen to diverge at points $x_i$, but it is finite and positive everywhere else. 

\subsubsection{Vacuum initial data with cosmological expansion}
\label{sec:expanding}

Imposing time symmetry is the simplest, but not the only way, to obtain 
initial data complying with the condition discussed in Section~\ref{sec:vacuum}.
One may choose, for instance, to retain the terms containing $K$ and $A_{ab}$ in
equation (\ref{eq:CTTconstraints}) (or equivalently $\Theta$ and $\sigma_{ab}$ 
in (\ref{m_constr})). This enables the description of spacetimes with vanishing
spatial conformal curvature and an initial 
non-zero volume expansion, both key properties for an observationally-motivated 
cosmological model.

Allowing for a non-zero extrinsic curvature results in two major complications: (i)
the Hamiltonian constraint becomes nonlinear, and requires a numerical solution, and
(ii) the momentum constraint is not identically satisfied in general, and will have
to be explicitly solved. In the case of BHL, an initial-data construction in this 
class has been proposed in~\cite{Yoo:2012jz}, where the 
Lichnerowicz-York conformal-transverse-traceless 
scheme~\cite{lichnerowicz:1944,York:1971hw,York:1972sj,York:1973ia}
was used, with a conformally-flat metric and a piecewise constant 
trace of the extrinsic curvature as free data: 
\beq
\begin{split}
h_{ij} &= \psi^4 \delta_{ij} \\
K_{ij} &= \frac{K}{3} \delta_{ij} + A_{ij}
\end{split}
\eeq
with
\bea
K=K_{\rm c} \, T(r) &=& K_{\rm c} \left\{
  \begin{array}{ll}
  0 & \textrm{for } 0 \le r \le l \\
  \left(\frac{(r-l-\sigma)^6}{\sigma^6}-1\right)^6&\textrm{for } l \le r \le l + \sigma \\
  1 & \textrm{for } l + \sigma \le r
  \end{array}
\right.
\eea
where $K_{\rm c}$, $l$ and $\sigma$ are suitable parameters.
The system (\ref{eq:CTTconstraints}) is then solved numerically for $\psi$
and $A_{ij}$.

The core idea of this
method is that the initial data for a conformally-flat, periodic BHL 
can be represented as a periodic cube with a spherical region at its
center where the geometry is instantaneously static 
(so that the constraint equations are locally identical to those
describing a single Schwarzschild singularity in isotropic coordinates), 
surrounded (after a thin transition shell) by a uniformly expanding 
region stretching out to the cube faces.

\section{Lattice arrangements}
\label{sec:arr}
Given that the real universe contains trillions or more discrete sources in multi-scale configurations, we cannot hope to model the locally inhomogeneous structure of the universe without simplifying assumptions. In this section we describe how to impose a lattice structure in order to handle local inhomogeneities in a way which both corresponds to a homogeneous structure at large scales and that can also be used for efficient analytical and numerical calculations. 

Turning to cosmological considerations for the background, it is desirable to distribute the sources on the background in a configuration that is at least roughly uniform. 
The most natural construction for a lattice universe is therefore to use one of the regular tessellations of constant-curvature 3-spaces \cite{Coxeter:1948}. The possible regular lattices for different background curvatures are listed in Table \ref{table_regular_lattices}. 
%
\begin{table}[t!]
\begin{center}
\begin{tabular}{|c|c|c|c|}
\hline
   $\sgn(\threeR)$ & Schl\"afli symbol & \# cells  & Cell shape \\
\hline
 + &  \{333\} &    5 &  Tetrahedron  \\
 + &  \{433\} &    8 &  Cube  \\
 + &  \{334\} &   16 &  Tetrahedron  \\
 + &  \{343\} &   24 &  Octahedron   \\
 + &  \{533\} &  120 &  Dodecahedron  \\
 + &  \{335\} &  600 &  Tetrahedron  \\
 0 &  \{434\} &  $\infty$ &  Cube  \\
 - &  \{435\} &  $\infty$ &  Cube  \\
 - &  \{534\} &  $\infty$ &  Dodecahedron  \\
 - &  \{535\} &  $\infty$ &  Dodecahedron  \\
 - &  \{353\} &  $\infty$ &  Icosahedron   \\
\hline
\end{tabular}
\end{center}
\caption{The table shows the regular tessellations possible for 3-dimensional spaces of constant curvature $\threeR$ with $\sgn(\threeR)$ positive $(+)$, zero $(0)$ or negative $(-)$. The Schl\"afli symbol gives the lattice structure $\{p,q,r\}$ where $p$ is a property of a face, $q$ is a property of a polyhedron cell and $r$ is a property of the lattice itself. Specifically, $p$ is the number of edges of the face, $q$ is the number of faces meeting at the corner of the polyhedron cell and $r$ is the number of cells meeting along an edge in the lattice. The number of cells (\#) is given in the third column,  and the last column shows the polyhedron type of each the cells.}
\label{table_regular_lattices}
\end{table}

\subsection{Symmetry considerations}
\label{symm}
\begin{figure}[t]\label{coxeter_complex}
\begin{centering}
\includegraphics[width=6.5cm]{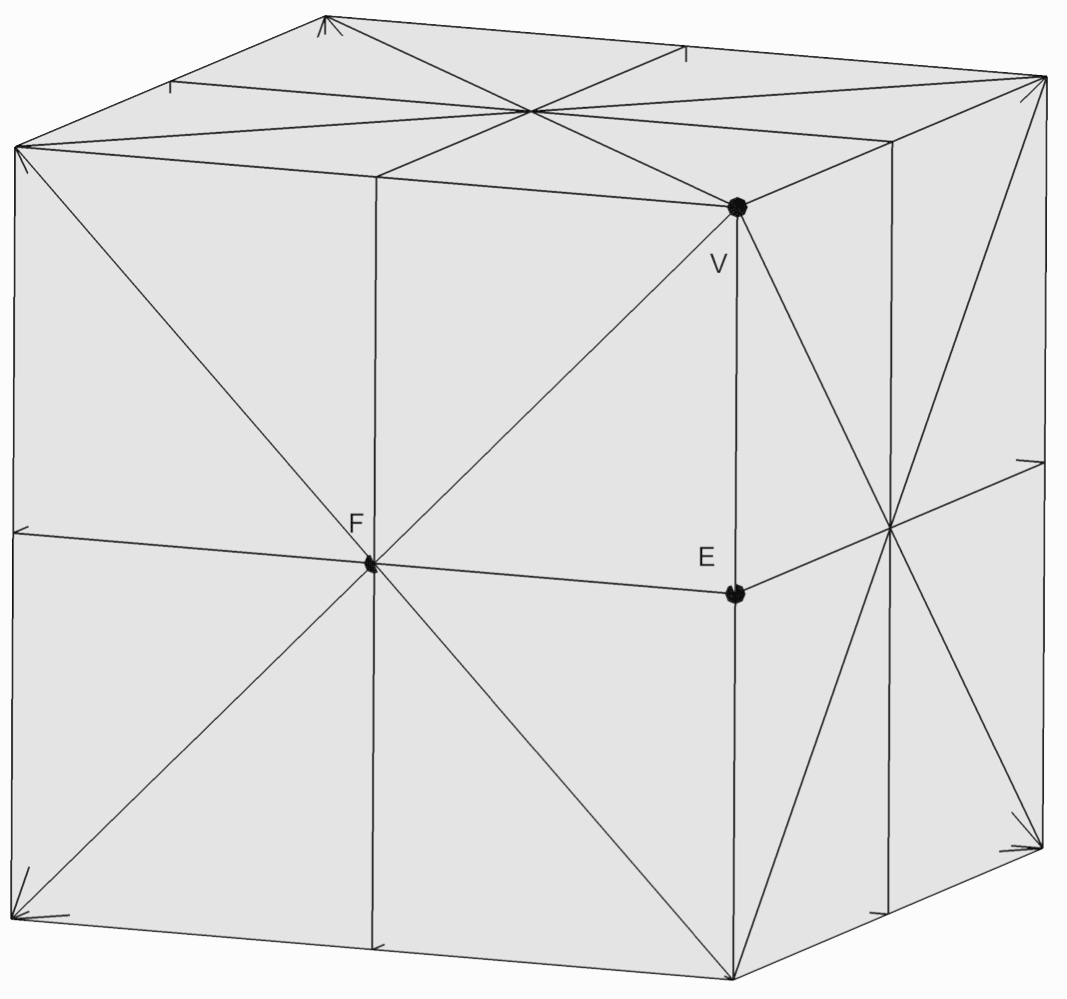} \qquad
\includegraphics[width=6.5cm]{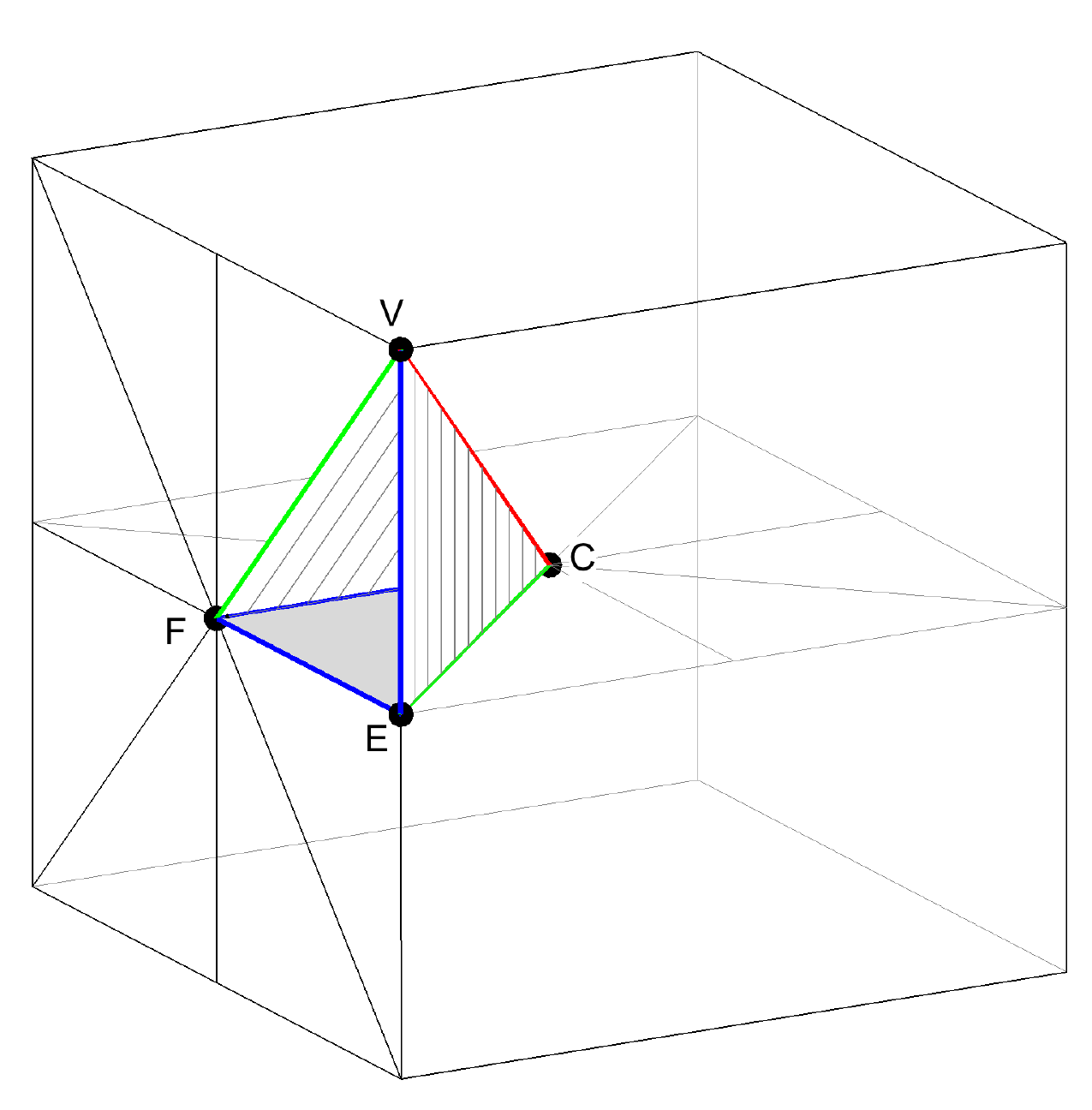} 
\par\end{centering}
\caption{The left panel shows the all intersections between mirrors and faces of a cubic cell.  A chamber in the cubic cell is displayed in the right panel (see text). The six types of mirror intersections are also shown, FC with $s=4$, VE and VC with $s=3$ and EF, EC and VF with $s=2$, where $s$ is the number of intersecting mirrors. Reproduced from~\cite{Clifton:2014lha}.}
\centering
\end{figure}

By construction, a regular lattice structure is endowed with a number of symmetries. If the background geometry has isometries that are hypersurface orthogonal Killing vector fields, then any hypersurface orthogonal to the Killing field also defines a reflection symmetry, at least in a neighbourhood of the surface. While a lattice structure has no continuous symmetries, it can inherit some of those reflection symmetries, which by nature are discrete. In a regular lattice (see Table \ref{table_regular_lattices}) each face in the lattice is part of a reflection symmetry surface (or \emph{mirror}). Therefore the edges in the lattice are curves that are intersections of three or more symmetry surfaces. It follows that the edges are curves with local rotational symmetry (LRS, Section~\ref{field_restrictions}). The lattice vertices enjoy an even higher degree of symmetry being meeting points of several edges, and are in fact locally isotropic. These kinds of discrete symmetries imply certain restrictions on the dynamics at edges and corners that will be discussed in the following. Besides edges, there are other curves which are also intersections of mirrors. Their meeting points also have higher symmetries. Referring to Figure~\ref{coxeter_complex}, the points with higher symmetry, in addition to vertices (V), are the midpoints of edges (E), faces (F) and cells (C) respectively. 
%
The set of all mirrors subdivides the lattice in tetrahedral regions called chambers \cite{Borovik_AV&A:2010}, as illustrated in the case of a cubic lattice in Figure~\ref{coxeter_complex}. The chambers are all identical up to reflections and translations. 

Consider now a matter configuration that respects all the reflection symmetries of the lattice. An example of such a configuration is obtained by placing identical sources at all cell centres. 
By construction, all physical fields associated with the sources must then respect all reflection symmetries leading to certain restrictions on those fields. This applies, in particular, to the kinematic quantities as well as the curvature. If a reflection symmetry acts in an initial data hypersurface, then it must be preserved by Einstein's evolution equations. This follows as the same data is evolved on both sides of the reflection surface.

\subsubsection{The 2+1+1 decomposition.}
\label{decomps}

The relevant fields for cosmology include (polar) scalars $S$, spatial (polar) vectors $S_a$ and second rank spatial (polar) tensors $S_{ab}$, as well as pseudo-scalars (or axial scalar), pseudo-vectors (or axial vector) and pseudo-tensors. Always present in vacuum are the kinematic fields, i.e.\ the expansion ($\Theta$) and the shear ($\sigma_{ab}$), as well as the gravitoelectric ($E_{ab}$) and gravitomagnetic ($H_{ab}$) curvatures. The field equations in the form given in \cite{Maartens:1997} also contain  the differentiated fields $D_a\Theta$, $(\div S)_a$ and $(\curl\! S)_{ab}$ where $S_{ab}$ represents any of the fields $\sigma_{ab}$, $E_{ab}$ and $H_{ab}$. The relevant field types are therefore scalars, spatial vectors, and spatial symmetric trace-free second rank tensors.

The discrete symmetries present in a lattice lead to restrictions on the kinematic and field variables on a mirror surface $\Mscr$. To describe these restrictions it is helpful to employ a 2+1+1 decomposition adapted to $\Mscr$ 
\cite{Clifton:2016mxx}.
This can be achieved by defining a projection tensor defined by
\begin{equation}\label{mirror_projection}
   f^a{}_b = h^a{}_b - n^a n_b
\end{equation}
where $n^a$ is a spatial unit vector field defined on the initial hypersurface.  If $n^a$ is normal to $\Mscr$, then a spatial vector field $V^a$ can be projected onto $\Mscr$ by defining a transverse 2-vector part perpendicular to $n^a$ by $\fperp{V^a}= f^a{}_b V^b$. The remaining part is parallel to $n^a$ and is given by $\fpar{V^a} = \fpar{V} n^a$ where $\fpar{V} =V^a n_a$. The spatial vector can therefore be irreducibly decomposed as
\begin{equation}
   V^a = \fpar{V^a} + \fperp{V^a} \, .
\end{equation}
Similarly, a symmetric trace-free spatial tensor $S_{ab}$ can be irreducibly decomposed into scalar, transverse 2-vector and symmetric trace-free transverse 2-tensor parts by
\begin{equation}
   S_{ab} = \fscalar{S_{ab}} + \fvec{S_{ab}} + \ftensor{S_{ab}}
\end{equation}
where
\begin{equation}\label{211fields}
   \fscalar{S_{ab}} = \fscalar{S}(n_a n_b - \tfrac12 f_{ab}) \, , \qquad
      \fvec{S_{ab}} = 2n_{(a} \fvec{S_{b)}} \, , \qquad
   \ftensor{S_{ab}} = \bigl(f_{(a}{}^c f_{b)}{}^d 
                     -\tfrac12 f_{ab} f^{cd} \half\bigr)S_{cd} \, ,
\end{equation}
with the scalar and vector parts given by $\fscalar{S} =n^a n^b S_{ab}$ and $\fvec{S_a = f_a{}^b n^c S_{bc}}$, respectively. 

\subsubsection{Field restrictions on symmetry-invariant subspaces.}
\label{field_restrictions}
In this section we discuss symmetry restrictions in the tangent spaces of invariant subspaces formed by the discrete symmetry groups of lattices. The reader may consult references \cite{Clifton:2013jpa}, \cite{Korzynski:2015isa} and \cite{Clifton:2016mxx} for a more detailed treatment. For a 3-dimensional lattice, the subspaces can have dimension 0, 1 or 2. The zero-dimensional case occurs when three or more mirror surfaces meet at a point, and their normals span the tangent space at that point. The one-dimensional case is when two or more symmetry surfaces meet along a common curve while the two-dimensional case corresponds to the subspace of a single mirror surface.

If a reflection symmetry leaves an initial 3-metric invariant, then it will continue to be a symmetry, at least in its future Cauchy development. Therefore, any restriction on a tensor implied by the reflection symmetry will continue to be valid at least until the appearance of a Cauchy horizon. It is therefore important to understand the restrictions imposed by the symmetry on the initial hypersurface. Given that a reflection symmetry is an improper Lorentz transformation, the restrictions will depend on the parity property of the relevant tensor quantities, i.e.\ whether they are polar or axial tensors. Choosing $n^a$ in \eqref{mirror_projection} to be normal to the mirror, the scalars involved are the ordinary (polar) scalars $\Theta$, $\fscalar{\sigma}$ and $\fscalar{E}$ and the axial scalars $\fscalar{\half(\curlsigma)}$, $\fscalar{\half(\curlE)}$ and $\fscalar{H}$. The first three are polar since there is no Levi-Civita tensor in their definition and only even multiples of  $n^a$.

\paragraph*{Field restrictions on a mirror surface.}

Since axial scalars change sign under reflection they must all vanish on a mirror surface. This gives the restrictions $\fscalar{(\curl\sigma)} =0$, $\fscalar{(\curlE)} =0$ and $\fscalar{H} =0$. Therefore, the surviving scalars are $\Theta$, $\fscalar{\sigma}$, $\fscalar{E}$ and $\fscalar{(\curlH)}$. Related arguments for the vectorial and tensorial parts reveal that the surviving parts are the axial vectors $\fvec{(\curl\sigma)}_a$, $\fvec{(\curlE)}_a$ and $\fvec{H}_a$ and the polar tensors $\ftensor{\sigma}_{ab}$, $\ftensor{E}_{ab}$ and $\ftensor{(\curlH)}_{ab}$.

\paragraph*{Field restrictions on the intersection of two mirror surfaces.}
This is the situation when there is a discrete symmetry group with two elements corresponding to a rotation by an angle $\pi$. It follows that the surfaces must be at right angles, relative to each other. The intersection of the surfaces defines a curve $\Cscr$. In this case, rather than picking a normal to one of the surfaces, it is more natural to align the vector $n^a$ with $\Cscr$. This implies in particular that the 2-plane in the 2+1+1 decomposition will now be perpendicular to the two symmetry surfaces. It is important to keep in mind that the 2+1+1 notations $\fscalar{S}$, $\fvec{S}^a$ and $\ftensor{S}_{ab}$ therefore have a different meaning with respect to the symmetry surfaces.

In this case, all axial scalars must also vanish by the same argument and the remaining  scalars are $\Theta$, $\fscalar{\sigma}$, $\fscalar{E}$ and $\fscalar{(\curlH)}$. It can also be shown that all vector and axial vector parts must vanish. To give the restrictions on the tensor part we use a frame $(e_1 = n, e_2, e_3)$ where $e_2$ and $e_3$ are unit vectors orthogonal to the curves $\Cscr$. Note also that the tensor part, being symmetric and tracefree, has the two independent components $\ftensor{S}_{22}$ and $\ftensor{S}_{23}$. It can be shown that the remaining unrestricted parts are the polar tensor components $\ftensor{\sigma}_{22}$, $\ftensor{E}_{22}$, $\ftensor{(\curlH)}_{22}$ and the axial tensor components $\ftensor{(\curlsigma)}_{23}$, $\ftensor{(\curlE)}_{23}$, $\ftensor{H}_{23}$.

\paragraph*{Field restrictions on an LRS curve.}
This case has a special importance since it applies to special curves, such as lattice edges, which are relevant for analysing the large scale structure of the cosmological lattices (see Section~\ref{S3lattices} where  this is discussed using numerical and analytical approaches). When three or more mirror surfaces intersect along a curve $\mathcal{C}$, then the tangent space at a point on $\mathcal{C}$ has a continuous rotational symmetry group in the plane perpendicular to $\mathcal{C}$. This is what defines an LRS curve.  
Due to the continuous rotational symmetry, there can be no preferred vector or direction in that plane.
 Taking again the unit vector $n^a$ along $\mathcal{C}$ as base for the 2+1+1 decomposition, this implies that all vector and tensor parts must vanish. Since the axial scalars must again vanish, only the polar scalars $\Theta$, $\fscalar{\sigma}$, $\fscalar{E}$ and $\fscalar{(\curlH)}$ can be non-zero.

We can then write all the second rank fields on $\mathcal{C}$ in the form 
$S_{ab} = \fscalar{S}\half k_{ab}$ as given in \eqref{211fields} where $k_{ab}$ depends only on $g_{ab}$,
$u^a$ and $n^a$.
To write down the reduced field equations one would therefore generally speaking need to consider $n^a$ as a vector field in a neighbourhood of $\mathcal{C}$. 
However, on $\mathcal{C}$, $n^a$ would then be parallel propagated (\emph{i.e}.\ $\dot n^a =0$) by its definition as a unit vector on $\mathcal{C}$.
This leads to the reduced evolution equations for the remaining  variables \cite{Clifton:2014lha} 
\begin{equation}
 \begin{split}
             \dot\Theta &= 
                    -\tfrac13 \Theta^2 - \tfrac32 (\fscalar\sigma)^2 \\[3pt]
   \fscalar{\dot\sigma} &= -\tfrac23 \Theta\half\quart\fscalar\sigma
                           -\tfrac12(\fscalar\sigma)^2 - \fscalar{E} \\[3pt]
       \fscalar{\dot E} &= -\Theta\half\quart\fscalar{E}
                           +\tfrac32\half\fscalar\sigma\half\quart\fscalar{E}
                           +\fscalar{(\curlH)} \ .
 \end{split}
\end{equation}
Here $\fscalar{(\curlH)}$ acts as a driving term in the equation for $\fscalar{E}$ even though $H_{ab}$ itself is zero on $\mathcal{C}$ during the evolution. To do the full integration one therefore needs also the evolution of $H_{ab}$ in a neighbourhood of $\mathcal{C}$. However, it turns out that one can evaluate $\fscalar{(\curlH)}$ recursively as a power series in $t$ using the full evolution equations~\cite{Clifton:2014lha}. This does become cumbersome to do even for the first few terms though. We note that the role of $\curlH_{ab}$ as a driving term and the possibility of recursive evaluation is actually true even for the full system \eqref{m_ev}. In this sense, the Einstein equations could be considered as being ``almost ODEs".  

\paragraph*{Field restrictions at vertices of a regular lattice.}
In the case of a regular lattice, there are three types of LRS curves corresponding to the lines VE, FC and VC in Figure~\ref{coxeter_complex} (\emph{cf.~} Table 3 of \cite{Clifton:2014lha}).
This includes the edges, of which there are three or more meeting at each vertex. As has been noted above, the only nonvanishing parts of symmetric tracefree tensors $S_{ab}$ on LRS curves are polar scalars of the type $\fscalar{S}_{ab} = S_{ab} n^a n^b$ where $n^a$ is directed along the curve. At a vertex, this applies to each of three independent directions implying that these must all vanish. It follows that the tensors $\sigma_{ab}$, $E_{ab}$ and $H_{ab}$ are all zero at every vertex. It then also follows from the evolution equations \eqref{m_ev} that $\Theta$ is identically zero for all times at every vertex. We conclude that the vertices are a kind of equilibrium point, with vanishing Riemann tensor throughout their evolution.

\subsubsection{Other configuration issues}
\label{conf_issues}

\paragraph*{Configurations that minimize the potential energy.}
Besides the highly symmetric configurations discussed so far, it is also of interest to consider other types of configurations. An obvious reason is that the  universe is certainly not a regular lattice. One issue relevant to the real universe is that of interaction energies between masses, and how they relate to the matter distribution. The interaction energies depend on the distances between objects, which will of course be different depending on the degree of clustering of the masses. If there is only one object in the universe, that distance is zero. It could be argued that this would be the situation if a Schwarzschild black hole was formed by a collapse of a spherically symmetric matter configuration. Then the interaction energy could be taken as infinite since there would be no way even in principle of bringing the matter back to infinity. At the other extreme, the sources could be placed in a configuration that maximizes the least distance between any two of the sources.\footnote{For the 2-sphere, this is known as the Fejes Toth problem \cite{Whyte:1952}.} One might at first think that the regular configurations should meet this requirement. However, this is not always the case as can be readily understood. To see this, consider the case of a regular tessellation of the 2-sphere. Inscribing a  cube in the sphere gives such a configuration with eight sources corresponding to the eight vertices of the cube. Assuming that the cube is in an upright position, let the four lower vertices of the cube be fixed and rotate the upper vertices by 45 degrees in a horizontal plane. This gives a configuration that is a semiregular polyhedron known as a square antiprism. It is obvious that it has a larger mean distance than the inscribed cube. One could further increase the mean distance by a deformation that makes the distance between all neighbouring points equal while still keeping the square antiprism form.
The resulting configuration actually maximizes the least distance for eight points on a 2-sphere \cite[p.2781]{Weisstein:2003}\cite{Whyte:1952}. Interestingly, this form is realised in nature in some molecules. For example, the most stable allotrope of sulfur, $S_8$, has this form.
 
The same configuration also minimizes the potential energy for eight identical sources with a $1/r$ interaction confined on a 2-sphere. Finding the configuration that minimizes the energy for such an interaction is known as the Thomson problem~\cite{Ashby&Brittin:1986}. In general, a configuration which solves the Fejes Toth problem does not minimize the energy. One example is the case of seven sources that is solved by different configurations for the two problems~\cite{Whyte:1952}. While this discussion so far has concerned only lower-dimensional configurations with few sources, it has a wider interest when considering several sources in the context of the structures we see in the universe. In particular, the observed filaments and voids form a cell-like structure, much like that used in discrete cosmology (\emph{cf}.\ discussion in Section~\ref{sec:struct} on structuration).

\paragraph*{Configurations with sources in randomised positions.}
Up to the present time, most work on explicit models in discrete cosmology has focused on configurations with a high degree of regularity (see however Section~\ref{sec:struct} and references therein for a discussion of general analytical results). 
The analytical results obtained for the regular models depend on the existence of LRS curves. It is then natural to ask if one could decrease the regularity while still keeping some LRS curves. This can actually be done in a few different ways. As mentioned above, the existence of an LRS curve relies in turn on the presence of at least three reflection symmetries meeting along a common curve. Consider now a model that is invariant under a reflection group, and having cells with faces that lie in reflection invariant surfaces. If we distribute sources in one of the cells (the seed cell) in a completely random manner, then we can spread that distribution to all the other cells by appropriate reflections. To do this we should reflect the distribution in one of the faces of the seed cell. Then one of the cells neighbouring the seed cell will be filled by a mirror image of the original distribution, {\it etc}. When continuing this process, some cells will contain the original distribution and some will contain its mirror image. In order not to destroy the symmetry group, all faces must have  the original distribution on one side and the mirror image on the other side. It is then clear that for this to work, there must be an even number of cells meeting at each edge. 

The only regular lattices that meet this requirement are the hyperspherical 16-cell, the flat cubic lattice and the dodecahedron hyperbolic lattice, all of which have four cells meeting at the edges (see Table~\ref{table_regular_lattices}). Another possible strategy is to use a chamber as seed cell. Then any of the regular lattices can be used since the chambers are themselves mirror images of all their neighbours. One could then argue that the most random configurations would be the ones which contain the least number of chambers. For the 3-sphere background this would be the 5-cell, which contains 120 chambers.

\subsection{Lattices with extra fields}
\label{ssec_extra}

The six perfectly regular tessellations of the 3-sphere discussed above have also been considered in the presence of a non-vanishing cosmological constant \cite{Yoo:2014boa,Durk:2016yja}, electric charges \cite{Bibi:2017urt}, and scalar fields \cite{Clifton:2017hvg}. We will discuss these additional configurations in this section.

\subsubsection{Lattices with $\Lambda$.}
\label{sssec_lambda}

In the presence of non-zero $\Lambda$ it is possible to find exact initial data when the initial hypersurface is no longer time-symmetric, but instead has constant mean curvature. That is, if we make the choice that the trace and trace-free parts of the extrinsic curvature are given by

\beq \label{sdsk}
K^2 = 3 \Lambda \qquad {\rm and} \qquad A_{ij}=0,
\eeq
then the momentum contraint is solved immediately, and the Hamiltonian constraint again reduces to the form given in equation (\ref{genH}) \cite{Durk:2016yja}.

In this case, the relevant geometry to consider in order to find superposable solutions is Schwarzschild-de Sitter. In an isotropic coordinate system it is possible to find a foliation of this spacetime that contains a leaf with the extrinsic curvature specified by equation (\ref{sdsk}), and with intrinsic geometry given by \cite{Nakao:1990gw}
\beq
g_{\rm SdS} = \psi_+^4 ( d \tilde{r}^2 + \tilde{r}^2 d \Omega^2 ) \, ,
\eeq
which can be seen to be the same expression as occurs for the Schwarzschild geometry in the absence of $\Lambda$. Thus, in this case, one can conclude that the intrinsic geometry for an initial hypersurface with constant mean curvature in the presence of non-zero $\Lambda$ are exactly the same as the time-symmetric surfaces in the absence of $\Lambda$.

In contrast to the case where $\Lambda=0$, we can specify whether the initial hypersurface is expanding ($\Theta > 0$) or contracting ($\Theta < 0$). This changes the structure of the apparent horizons, and in particular whether a given horizon is a marginally inner or outer trapped. Further, as we are considering a superposition of Schwarzschild-de Sitter geometries, we can also have cosmological as well as black hole horizons. 

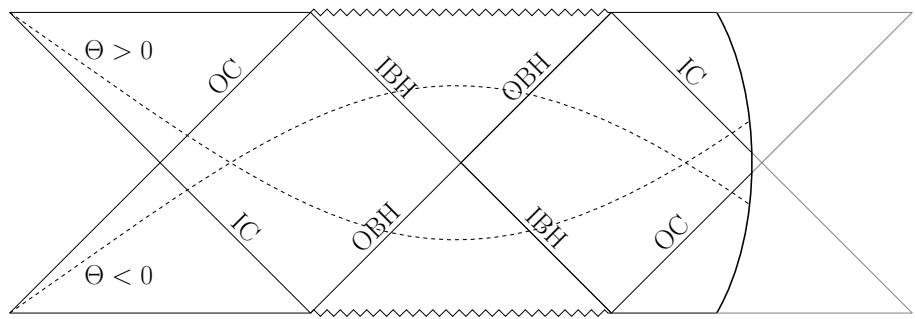
\begin{figure}[tbp]
\scalebox{0.5}{
\begin{tikzpicture}
\node (I)    at ( 4,0)   {};
\node (1)    at ( 7.8,1.2)   {};
\node (2)    at ( 7.8,-1.2)   {};
\node (II)   at (-4,0)   {};
\node (III)  at (0, 2.5) {};
\node (IV)   at (0,-2.5) {};
\node (V)   at (-12,0) {};
\node (VI)   at (12,0) {};
\node (VII)   at (-8,-2.5) {};
\node (3)   at (-9.1,3){$\Theta > 0$};
\node (4)   at (-9.1,-3) {$\Theta < 0$};
\node (VIII)   at (-8,2.5) {};
\node (IX)   at (8,-2.5) {};
\node (X)   at (8,2.5) {};

\path  
  (II) +(90:4)  coordinate  (IItop)
       +(-90:4) coordinate (IIbot)
       +(0:4)   coordinate                  (IIright)
       +(180:4) coordinate (IIleft)
       ;
\draw (IIleft) -- 
         node[midway, above left]    {}
         node[midway, above, sloped] {OC}
     (IItop) --
         node[midway, above, sloped] {IBH}
      (IIright) -- 
          node[midway, above, sloped] {OBH}
     (IIbot) --
          node[midway, above, sloped] {IC}
          node[midway, below left]    {}    
      (IIleft) -- cycle;

\path 
   (I) +(90:4)  coordinate (Itop)
       +(-90:4) coordinate (Ibot)
       +(180:4) coordinate (Ileft)
       +(0:4)   coordinate (Iright)
        ;

    \draw (Ibot) -- (Ileft)--(Itop); 
     \draw (Itop) -- 
     node[midway, above, sloped] {IC}
     ($(Itop)!.93!(Iright)$); 
         \draw (Ibot) --      
         node[midway, above, sloped] {OC}
         ($(Ibot)!.93!(Iright)$);
             \draw [gray]($(Itop)!.93!(Iright)$) -- (Iright); 
     \draw [gray]($(Ibot)!.93!(Iright)$) -- (Iright);

\draw  (Ileft) --
  node[midway, above, sloped] {OBH}
       (Itop);
       
     \draw  (Ileft) --
  node[midway, above, sloped] {IBH}
       (Ibot);
\path 
   (V) +(90:4)  coordinate (Vtop)
       +(-90:4) coordinate (Vbot)
       +(180:4) coordinate (Vleft)
       +(0:4)   coordinate (Vright)
        ;
\draw (Vtop) -- (Vright) -- (Vbot) ;
\path 
   (VI) +(90:4)  coordinate (VItop)
       +(-90:4) coordinate (VIbot)
       +(180:4) coordinate (VIleft)
       +(0:4)   coordinate (VIright)
        ;
\draw  [color=gray!100] (VItop) -- (VIleft) -- (VIbot) ;
\draw[decorate,decoration=zigzag](IItop) -- (Itop)
      node[midway, above, inner sep=2mm] {};

\draw[decorate,decoration=zigzag](IIbot) -- (Ibot)
      node[midway, below, inner sep=2mm] {};
      
\draw(Vtop) -- (IItop)
      node[midway, above, inner sep=2mm] {};

\draw(Vbot) -- (IIbot)
      node[midway, below, inner sep=2mm] {};
      
\draw(Itop)--($(VItop)!.65!(Itop)$)
      node[midway, above, inner sep=2mm] {};

\draw(Ibot) --($(VIbot)!.65!(Ibot)$)
      node[midway, below, inner sep=2mm] {};
      
\draw[thick, dashed] 
    (Vbot) to[out=35, in=145, looseness=1.3]
    (2);
    \draw[thick, dashed] 
    (Vtop) to[out=-35, in=-145, looseness=1.3]
    (1);

\draw[very thick]($(VItop)!.65!(Itop)$) to[out=-60, in=60, looseness=0.8]($(VIbot)!.65!(Ibot)$) ;
\draw  [color=gray!100] (VItop) -- ($(VItop)!.65!(Itop)$) ;
\draw  [color=gray!100] (VIbot) -- ($(VIbot)!.65!(Ibot)$)  ;
     
\end{tikzpicture}}
\caption{\label{fig:penrose} Penrose-Carter diagram for the region of spacetime around one of the black holes when $\Lambda \neq 0$. The horizons are labelled as stated in the text. Reproduced from \cite{Durk:2016yja}.}
\end{figure}

Figure \ref{fig:penrose} shows two example hypersurfaces (dotted lines) for two arbitrary values of $\Lambda$. In both cases there is an outer trapped cosmological horizon (OC), an outer trapped black hole horizon (OBH), an inner trapped black hole horizon (IBH) and an inner trapped cosmological horizon (IC), but the ordering of these surfaces changes depending on whether the spacetime is initially expanding or contracting. The solid curve on the right-hand side indicates a cut-off, beyond which the cosmology becomes more complicated (due to the presence of multiple other black holes).

It was found that distinct cosmological and black-hole horizons only exist for certain values of $\Lambda$, for each of the six regular lattice configurations that can tile the 3-sphere. This is an important condition for the existence of a multi-black hole cosmology, as a shared cosmological region outside each of the black holes is essential for the concept to make sense. In each case, this critical value of $\Lambda$ was found to be non-zero, and was observed to approach the Schwarzschild-de Sitter value of $M_0^{\,2} \Lambda = 1/9$ as the number of black holes was increased. It was also found that $\Omega_{\Lambda} \gtrsim \frac{4}{27} \Omega_m$ must be satisfied if the spacetime is to undergo eternal expansion, where $\Omega_\Lambda \equiv 3 \Lambda/\Theta^{\,2}$ and $\Omega_m \equiv 24\pi \rho/\Theta^{\,2}$.

Finally, it is also possible to numerically construct \Rthree lattice models with non-zero $\Lambda$: their initial-data construction and evolution were performed in \cite{Yoo:2014boa}, where the expansion history was found to be very close to FLRW models with dust and $\Lambda$.

\subsubsection{Lattices with electric charge.}
\label{sssec_electric}

Another interesting generalisation of the vacuum initial data is to investigate the effect of non-zero electric charge \cite{Bibi:2017urt}. For this case, a time-symmetric set of constraint equations for the Einstein-Maxwell system can be derived, and is given as follows:
\begin{eqnarray} \label{econ}
  \threeR = 2 E_i E^i \qquad {\rm and} \qquad  D_i E^i  = 0,
\end{eqnarray}
where $E_i$ are the spatial components of the electric field. By choosing 
\begin{eqnarray}
   ds^2 = \chi^2\psi^{2} (d\tilde{r}^2 + \sin^2 \tilde{r} \, d\Omega^2) \qquad {\rm and} \qquad
   E_i = \partial_i \ln[\chi/\psi],
\end{eqnarray}
it can be shown that the contraints in equation (\ref{econ}) are satisfied if
\beq
\bar\Delta^2 \psi = \frac{\bar \threeR}{8}\psi \qquad {\rm and} \qquad \bar\Delta^2 \chi = \frac{\bar \threeR}{8}\chi \, .
\eeq
As before, multi-black hole cosmologies can be constructed in this case by superimposing time-symmetric slices through the appropriate single-black-hole solution -- this time the Reissner-Nordstr\"{o}m solution. Analysis for lattices of electrically charged black holes can then be carried out, and it can be shown that the lattices are required in every case to have zero net charge (when summed over all black holes in the cosmological region). 

The particular case of a lattice of eight black holes in a hyper-cubic lattice was studied in \cite{Bibi:2017urt}, with black holes existing in antipodal pairs with equal and opposite charges to each other. These models were compared to the corresponding uncharged case, as well as to positively curved FLRW with the same total proper mass. It was found that as the black hole charges increased towards extremality, the discrepancy in the scale size of black-hole lattice models and FLRW models was decreased.


\subsubsection{Lattices with scalar fields.}
\label{sssec_scalar}

Black-hole lattices in the presence of a scalar field, $\varphi$, have been studied in \cite{Clifton:2017hvg}. In this case the condition of extrinsically flat initial data requires
\beq
\dot{\varphi} \, D_i \varphi =0 \, ,
\eeq
so that either $\dot{\varphi}=0$ or $D_i \varphi =0$. The first of these choices corresponds to the well-known Janis-Newman-Winicour solution \cite{PhysRevLett.20.878} (discovered previously by Fisher~\cite{Fisher:1948yn}, Bergman \& Leipnik \cite{PhysRev.107.1157}, and Buchdahl \cite{PhysRev.115.1325}). Although the most commonly used solution in the study of isolated masses, it was found that superpositions of these solutions at moments of time-symmetry did not offer viable cosmological models. This was due to the scalar field diverging, and the scale factor collapsing, in regions far from all of the mass points.

Instead, the condition $D_i \varphi =0$ offered plausible initial data for the type of cosmological models discussed in this review. These solutions were used in \cite{Clifton:2017hvg} to study the idea of black holes surviving through the minimum of non-singular cosmological ``bounce'', that itself was made possible by the existence of $\phi$. These models provide exact initial data for the evolution of primordial black holes in cosmology (the scalar field soon being forgotten, as the expansion rapidly reduces its density). They could also be used as initial data for a universe with non-zero energy-momentum at its maximum of expansion.

\section{Time evolution of a BHL}
\label{sec:evol}

An important property of BHL models is the behaviour of their large-scale expansion. The global evolution of members of the FLRW class has of course been well studied, leading to familiar relationships between topology and destiny in the zero-$\Lambda$ case, and to corresponding generalisations when $\Lambda \neq 0$. Depending on the values of its parameters, a model can expand forever or recollapse, have a Big Bang or a bounce, or even live in a quasi-static configuration akin to the Einstein static universe.

As one of the main goals of the study of BHLs is the construction of models that are homogeneous and isotropic above a certain scale, it is worth asking whether the large-scale behaviour of these models mirror that of the corresponding FLRW models qualitatively and quantitatively. As was described in Section~\ref{sec:arr}, properties such as the impossibility of a time-symmetric spatial hypersurface in models with zero or negative conformal spatial curvature and with $\Lambda=0$ are preserved, because they are a direct, universal consequence of the structure of Einstein's constraint equations. This property and its physical implications have also been discussed in the context of 
inhomogeneous and anisotropic inflation in~\cite{Kleban:2016sqm}.

More specific, quantitative statements about the sign and magnitude of the backreaction effects discussed in Section~\ref{sec:ave} requires a more detailed knowledge of the geometry of the system. Typically, the  description of the large-scale expansion of a BHL starts with the  definition of a reference subspace, whose proper extension is taken to represent a measure of scale for the cosmology. As the geometry of a BHL is inhomogeneous, different subspaces will have different properties, and particular care must therefore be exercised in their choice. Ideally, the length of a reference curve, for example, should reflect genuine large-scale properties of the system, and should be minimally affected by local inhomogeneities. For regular lattices, the edges of each constituent cell are special curves that satisfy these requirements, and have so far been the primary choice for most existing studies. The area of a cell's face is another possibility. Different measures of scale are 
discussed and compared in~\cite{Clifton:2012qh} and~\cite{Bentivegna:2012ei}.

Another critical aspect in the comparison of BHLs to FLRW models is the choice of a best-fit member of this class, an operation that is inherently ambiguous~\cite{Ellis:1987zz}. The possibilities largely
depend on the model at hand: spatially closed models, for instance, have a finite mass, which can be used to identify an FLRW counterpart for each BHL, as was done in~\cite{Clifton:2012qh}. If the BHL is conformal to the 3-sphere, curves in the two models can then be identified if they are described by the same coordinates on the conformal \Sthree. This procedure shows that if all the mass in a closed FLRW universe is concentrated in a few black holes according to the metric described in Section~\ref{sec:arr}, then the proper distance between points converges to the FRLW value as the number of black holes increases, as illustrated in Figure~\ref{fig:s3scale}.

\bfi
\center
\includegraphics[width=0.5\textwidth]{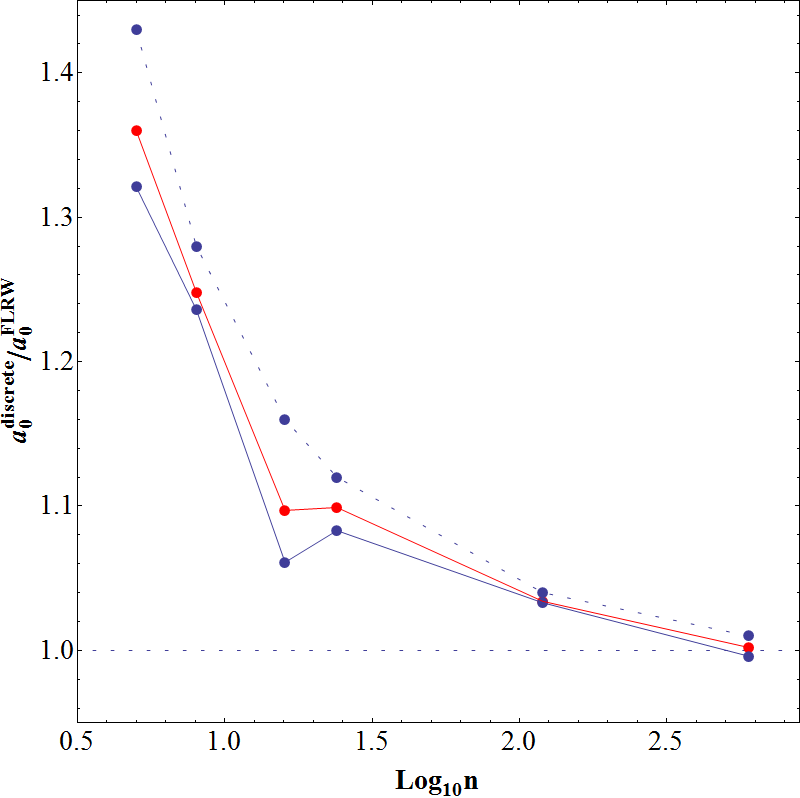}
\caption{The relative scale of a spherical BHL with respect to a closed
FLRW solution of the same total mass, as a function of the number of black
holes $n$. The two solid curves correspond two different definitions of scale, while the dashed curve represents the same quantity for a BHL constructed with the approximate Lindquist-Wheeler method. Reproduced from~\cite{Clifton:2012qh}.\label{fig:s3scale}}
\efi

The opposite procedure has also been explored, and can be used to 
define an effective mass for the BHL, as discussed in Section~\ref{sec:effect}.
In this case, one chooses the reference FLRW model as the one where curves of 
identical coordinate specification have the same proper length at a chosen time.
Fitting a FLRW model via the initial scale is particularly convenient 
if one is to compare the expansion of lengths as a function of time, 
because this prescription guarantees that the initial scale is the same
in the two models. For flat and open cosmologies, where the total mass
is not finite, other properties such as the initial expansion rate 
can be used to match FLRW and BHL cosmologies~\cite{Bentivegna:2013jta}.
The full evolution of a BHL has been carried out in two cases so far, models 
where spatial hypersurfaces are topological 3-spheres (henceforth 
referred to as \emph{\Sthree lattices}), and models where spatial 
hypersurfaces are topological three-tori (\emph{\Tthree lattices}).

\subsection{\Sthree lattices}\label{S3lattices}
This class of models was the one originally considered
by Lindquist and Wheeler for the application of the Schwarzschild-cell
method (where the evolution of a cell edge length obeys
the Friedmann equation). As these models admit time-symmetric spatial 
hypersurfaces where Einstein's constraints become linear, exact initial
data are possible~\cite{RevModPhys.29.432}. This initial data has recently been analysed in 
detail in \cite{springerlink:10.1007/BF01889418,Clifton:2012qh}. These
works paved the way for the first full numerical-relativity evolution
of the 8-black-hole model, carried out in~\cite{Bentivegna:2012ei}. This was
achieved by first showing that a coordinate transformation representing
a stereographic projection of the \Sthree lattice (suitably embedded in \Rfour)
onto \Rthree establishes a map between \Sthree lattices and the well-known 
Brill-Lindquist initial data class for multiple black holes~\cite{Brill:1963yv}. The 
simulations in~\cite{Bentivegna:2012ei} were then carried out by evolving
the dual Brill-Lindquist dataset with standard tools~\cite{Loffler:2011ay,
kranc,mclachlan,carpetweb,carpet}. This allowed the proper length of the lattice edges to be determined as a function of the proper time of the geodesic observers whose worldlines were orthogonal to the initial hypersurface, and resulted in the 
plot displayed in Figure~\ref{fig:s3length}.

\bfi
\center
\includegraphics[width=0.65\textwidth]{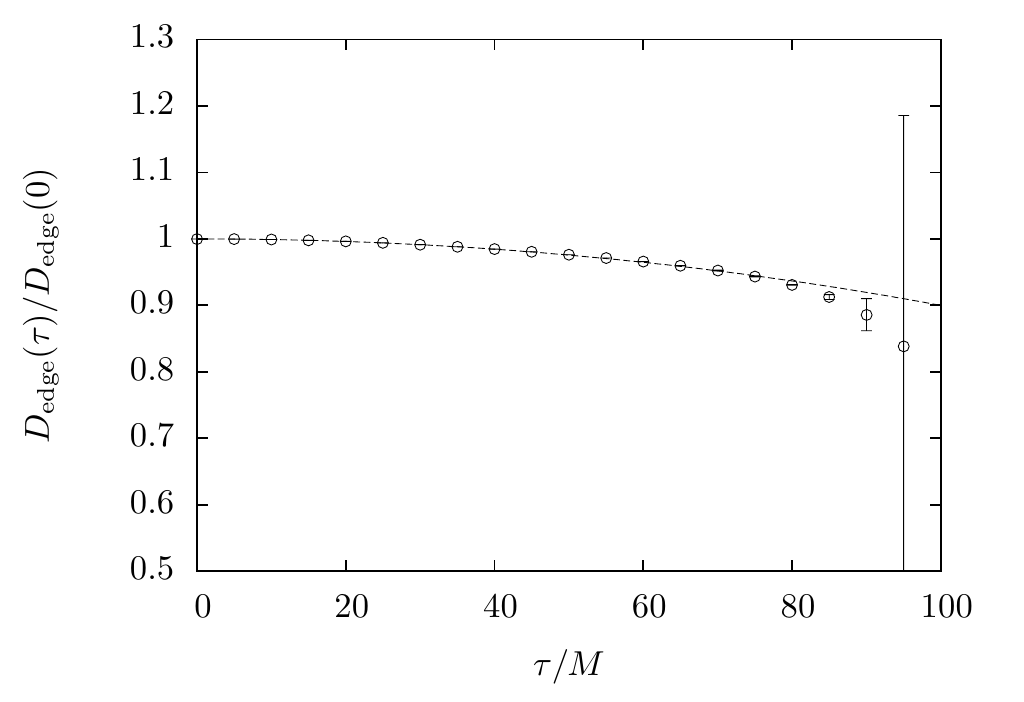}
\caption{
The evolution of the proper length of a lattice edge $D_{\rm edge}(\tau)$ 
as a function of the proper time $\tau$, for the initial geodesic normal observers. 
The dashed line represents the length scaling of a suitably-fitted, closed
FLRW model, which remains close to the lattice scaling  throughout the simulation (within the numerical error bars). As the lengths of the cell edges
contract, they become less and less resolved, leading to a growing numerical
error in the determination of $D_{\rm edge}$. Reproduced from~\cite{Bentivegna:2012ei}.
\label{fig:s3length}}
\efi

This result can be compared to the analytical approaches proposed in~\cite{Clifton:2016mxx,
Clifton:2013jpa,Korzynski:2015isa} and~\cite{Liu:2015bwa}. Here,
different approximate models of the \Sthree lattices are constructed, exploiting
useful cancellations due to the spatial symmetries or modelling the BHL
large-scale properties via representative flat \emph{skeletons} inspired by Regge
calculus~\cite{Regge:1961px}. As discussed in Section~\ref{field_restrictions}, in~\cite{Clifton:2016mxx,Clifton:2013jpa,Korzynski:2015isa} certain special
loci in the lattices, such as the cell edges, are studied in the orthonormal-frame
formalism~\cite{vanElst:1996dr}. This approach highlights the fact that the 
geometry of these curves is governed, to a very good degree of approximation,
by ODEs. Their evolution is therefore almost completely decoupled from the surrounding 
space, and can be obtained via a simple line integration, leading to Figure
\ref{fig:s3ode}.
\bfi
\center
\includegraphics[width=0.5\textwidth]{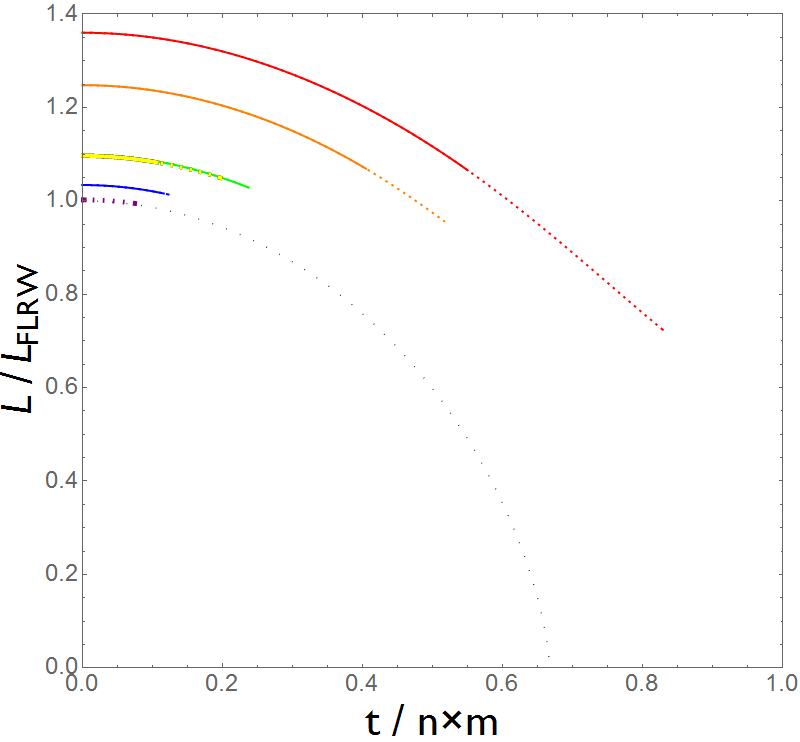}
\caption{
The evolution of the proper length of a lattice edge within the orthonormal-frame
approach. Reproduced from~\cite{Clifton:2013jpa}.
\label{fig:s3ode}}
\efi
The neglected term in the evolution ODEs, which can be quantified numerically~\cite{Korzynski:2015isa}
or via a Taylor expansion~\cite{Clifton:2016mxx}, is identically zero on the time-symmetric
hypersurfaces, but unfortunately grows substantially as one moves away from it, so that 
this method can only be used for part of the lattice evolution. 

The Regge-calculus approach, on the other hand, is based on the description of
a given spacetime via a piecewise-flat substitute, tessellated by polyhedra
glued together with certain deficit angles. The spacetime dynamics is then encoded
in these angles, and the evolution equations for the cell edges can be determined.
If a given, constant-time BHL hypersurface is approximated as a tessellation of 
identical regular polyhedra, this has the immediate advantage that the scaling
of lengths as a function of time has an obvious definition. Using this method,
Liu and Williams~\cite{Liu:2015bwa} were able to calculate the dependence between the volume of one
of the constituent polyhedra and its time derivative, and compare these quantities
against the closed FLRW models and the corresponding Lindquist-Wheeler lattices
with the same mass. The results are illustrated in Figure~\ref{fig:LW}. 

\bfi
\includegraphics[width=\textwidth]{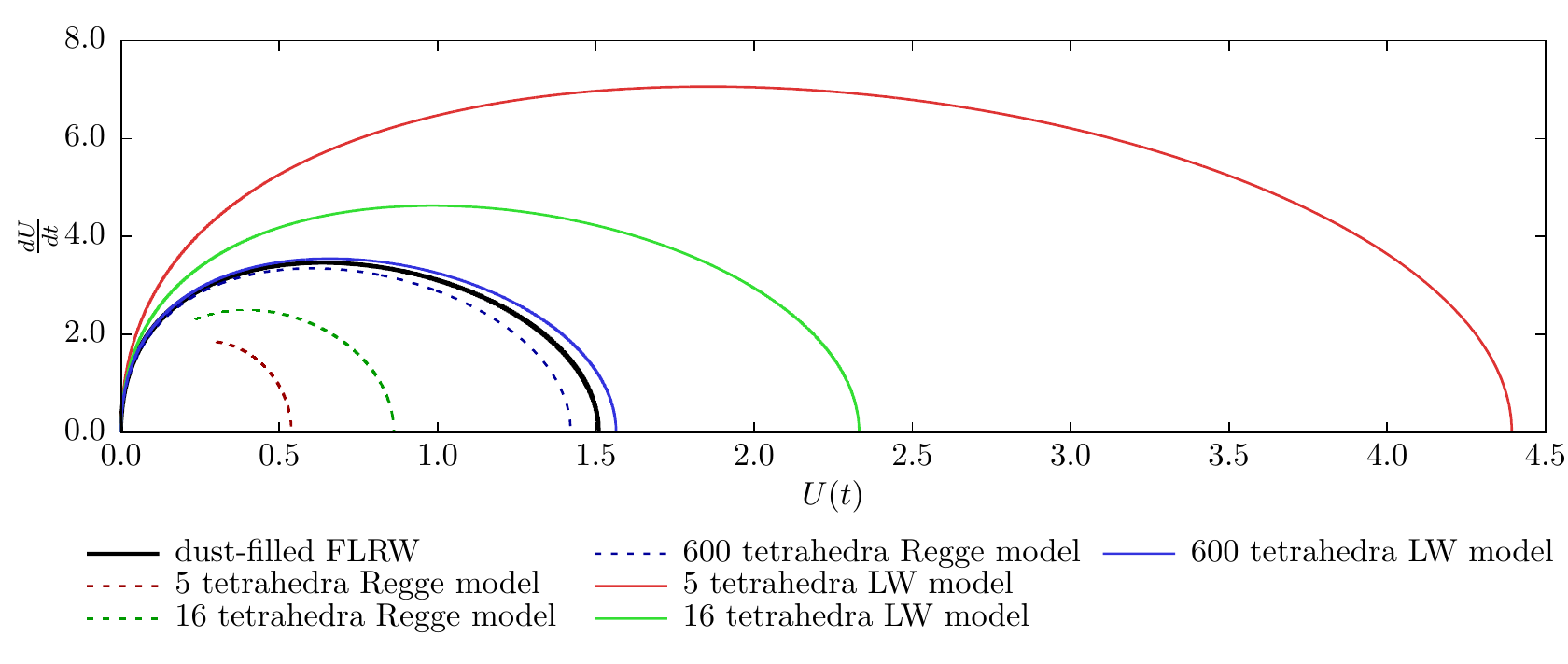}
\caption{
The evolution of a cell volume calculated using Regge calculus, compared to the corresponding
FLRW solution and to the Lindquist-Wheeler approximation. Reproduced from~\cite{Liu:2015bwa}.
\label{fig:LW}}
\efi

Two things can be noted from this result: first, the initial scale correction
predicted by the Regge-calculus method has opposite sign with respect to the
Schwarzschild-cell models and the exact initial data, a fact that the authors 
attribute to the excessively coarse-grained nature of their specific 
implementation~\cite{Liu:2015bwa}, and that could be reconciled with the other
methods by using more than one polyhedron to tessellate each lattice cell.
Second, constraints on the skeleton construction (such as the 
timelike nature of the edges connecting different, equal-time hypersurfaces)
prevent this model from stretching across the entire evolution of the lattice,
so that the spacetime regions away from the maximum-expansion hypersurface 
cannot be modelled.

\subsection{\Tthree lattices}
Models where spatial hypersurfaces are topological three-tori
certainly have a higher degree of cosmological relevance, but
the absence of conformal spatial curvature implies that no
time-symmetric hypersurfaces can exist in these spacetimes. This,
in turn, leads to nonlinear Einstein's constraints that have to
be solved numerically -- i.e., no exact initial data are known in
this class.

As discussed in Section~\ref{sec:expanding}, an initial-data construction
exists for this case, which has proven robust from the numerical
standpoint, leading to an elliptic PDE system which can be readily
solved with standard solution methods for elliptic problems~\cite{Yoo:2012jz,
Bentivegna:2013jta}, as long as the lattice compactness $\mu$, defined as
\bea
 \mu = \frac{M}{L} \label{eq:mudef}
\eea
with $M$ being the central mass and $L$ the edge length, is not too large.
 
This initial dataset was subsequently developed in time using full numerical
relativity~\cite{Yoo:2013yea,Bentivegna:2013jta}, for different values of the $M/L$ ratio. Not unlike the
\Sthree lattices, the length scaling in these models is also found to track 
that of a suitably fitted FLRW cosmology quite closely. In this case, 
the reference homogeneous and isotropic models are flat FLRW. The absence
of spatial curvature requires a change in the fitting procedure, which 
in this case is based on the initial length of a cube edge and its initial
time derivative (related to the initial choice of the trace of the extrinsic
curvature). This fit selects FLRW counterparts where the expansion of space
remains very close to the corresponding BHL, except for BHLs with high 
compactness, as illustrated in Figure~\ref{fig:t3length}.

\bfi
\center
\includegraphics[width=\textwidth]{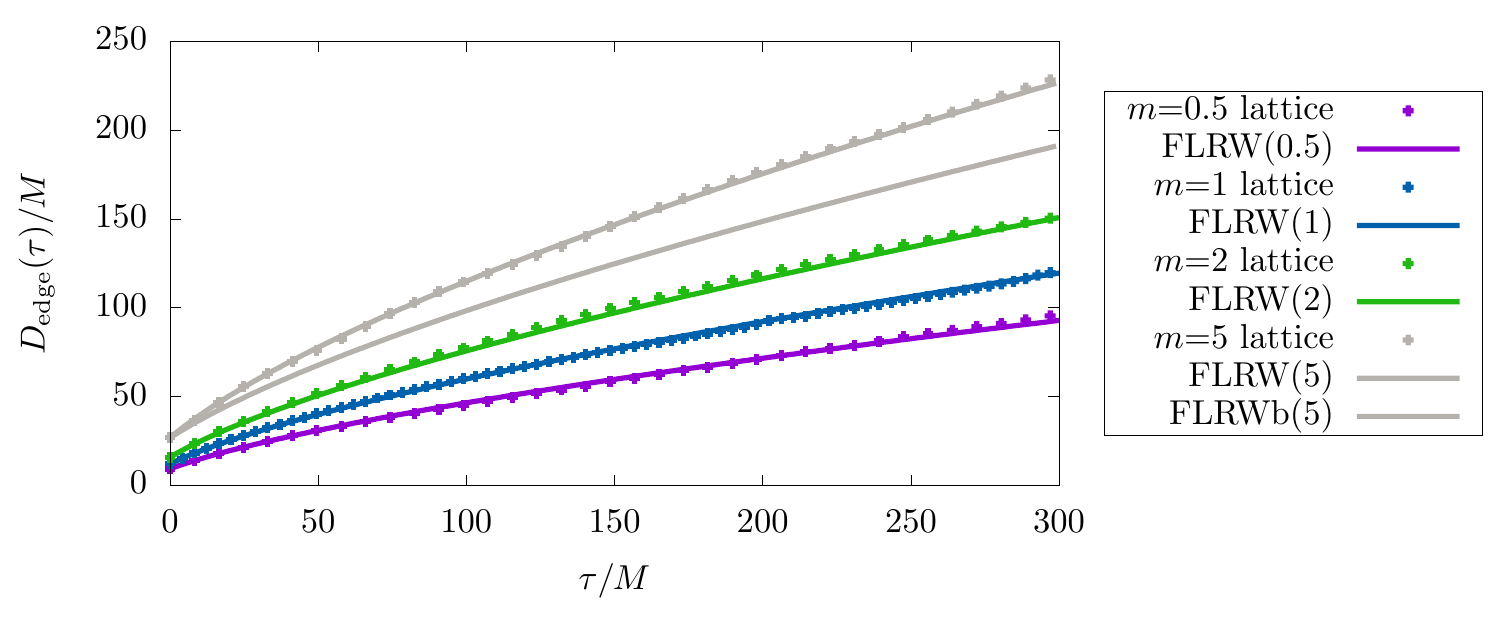}
\caption{
Comparison between the expansion of edge lengths in four \Tthree lattices, and in
the corresponding fiducial FLRW models. Reproduced from~\cite{Bentivegna:2013jta}.
\label{fig:t3length}}
\efi

As explained in Section~\ref{sec:gw}, this has more to do with the process
of fitting an FLRW model to the initial geometry of a high-compactness BHL than
with any qualitative change in the BHL dynamics as $M/L$ grows. This arises from the fact that an initially conformally-flat geometry leads
to the generation of spurious oscillatory modes in the gravitational field
at the beginning of the simulation (an effect also observed in 
conformally-flat initial data for binaries of compact objects, and dubbed
\emph{junk radiation} in that context). These modes affect lattices with large $M/L$ 
to a greater extent, and in particular obscure the leading-order
behaviour of the edge length, compromising the accuracy of the FLRW fit. The same construction can be replicated for models with a non-zero 
cosmological constant~\cite{Yoo:2014boa}. Similarly to the previous
cases, a good fit to an FLRW model can be found.

\subsection{Gauge}
A legitimate comparison of the evolution of edge lengths between
BHLs and the FLRW class requires a choice of time and space coordinates. 
In most cases, the obvious strategy is to measure the proper length of a curve as
a function of the proper time of the observers who are comoving with it.
This operation is trivial in FLRW models. In numerically-generated BHL
spacetimes, however, the numerical coordinates will usually not coincide with
the synchronous comoving gauge, as the latter is known to become pathological
in simulations of spacetimes containing black holes or other collapsed
regions (see, e.g., section 4.2.1 of~\cite{alcubierre2008introduction}).

A numerical coordinate transformation is therefore necessary in the postprocessing
phase, at least along the 1+1-dimensional subspaces spanned by the curve of interest. 
First, the constant-proper-time lines are solved for using
\beq
\tau(t,x) =\int_0^t \alpha(t',x) {\rm d} t'\, ,
\eeq
and the coordinate of comoving observers found via
\beq
x_{\rm g}(t,x_{\rm init}) = x_{\rm g}(t-\Delta t,x_{\rm init})-\int^t_{t-\Delta t} \beta^x(t',x_{\rm g}(t-\Delta t,x_{\rm init})) {\rm d} t' \, ,
\eeq
where the $x$ coordinate parametrises the curve on the constant-$\tau$ hyperspaces.
Finally, the curve's proper length as a function of $\tau$ is computed using
\beq
D(\tau)=\int_{\gamma_\tau} \left [  (-\alpha^2(\tau,\ell)+\beta^2(\tau,\ell)) (\partial_\ell t)^2 
                                   + \beta_i(\tau,\ell) \partial_\ell t \partial_\ell x^i 
                                   + \gamma_{ij}(\tau,\ell) \partial_\ell x^i \partial_\ell x^j \right ]^{1/2} {\rm d} \ell \, .
\eeq
An alternative is to use a geodesic integrator to track Gaussian observers (and 
compute their four-metric) along the chosen curve during the numerical evolution, 
and use this data to construct $D(\tau)$ directly in the appropriate gauge.

\subsection{Radiative components}
\label{sec:gw}
The BHLs explored to date are all obtained through the evolution of 
initial data that is conformally flat. Similarly to binary-black-hole
initial data, this choice results in a somewhat unphysical geometry,
which instantaneously lacks the tidal deformations induced by each of the 
black holes on every other one. Such a situation can only be achieved as a 
snapshot of a system of distorted black holes in the process of 
radiating away all non-stationary deformations. These oscillations
will then appear in the subsequent evolution as bursts of spurious
gravitational radiation moving away from the black holes. Unlike
the spurious radiation arising from asymptotically-flat binary-black-hole 
initial data, however, radiative modes cannot travel to infinity in the
compact (or periodic) spacetimes that are represented in BHLs. Aside from the generic attenuation
expected in an expanding geometry, and the possibility of reabsorption
by other black holes, they continue travelling across the lattice forever.

The radiative part of the gravitational field affects different observables in different ways, and to differing extents. The proper
edge length is the least affected, while higher-order quantities 
such as Einstein's constraint violations, the effective mass or
pressure, or the optical properties exhibit more significant oscillations~\cite{Bentivegna:2013jta,Bentivegna:2016fls}.
Section~\ref{sec:opt} will discuss a concrete example of this phenomenon
by showing the oscillations in the photon redshift (and therefore in
the redshift-distance relationship) along particular geodesics. 
Low-amplitude oscillations can also be picked out in the expansion
curve of the $m=5$ lattice in Figure~\ref{fig:t3length}, 
and in many of the effective quantities discussed in~\cite{Bentivegna:2013jta}.
Understandably, lattices of higher compactness are more affected, and effectively obstruct the fitting procedure
for the highest-mass case presented in~\cite{Bentivegna:2013jta}.

\section{Continuum limit and weak-field approximation}
\label{sec:cont}

\subsection{Continuum limit}\label{sec:continuumlimit}

Apart from the issues of time evolution and geometry, one of the more fundamental questions one might ask about black-hole lattices is the issue of the continuum limit, i.e.~the behaviour of the manifold when the 
mass of an individual black hole goes to zero, but the distance between the neighbouring ones shrinks so that the average mass density remains the same.

Consider a simple cubic lattice with a single black hole at the centre. Let $M$ be the black hole mass and $L$ the size of the lattice cell edge, as in Section~\ref{sec:evol}.
Disregarding any nonlinear effects of mass addition, the energy density of the model is 
\bea
 \rho = \frac{M}{L^3} \, .
\eea
We consider thus the limit in which $M\to 0$, while $\rho = \const$. 
This obviously corresponds to $L$ shrinking according to
the scaling law $L \propto M^{1/3}$.

Intuitively we may expect that the limit should correspond to an FLRW solution with dust, i.e. the Einstein-de Sitter universe, for the following two reasons:
\begin{enumerate}
\item the Schwarzschild radius $2M$ of the BHs becomes negligibly small in comparison with the Hubble scale of the underlying FLRW geometry, $H_0 \propto \rho^{-1/2}$, so that the influence of the granularity of
the mass distribution is likely to be negligible on cosmological scales;
\item the coordinate distance between neighbouring black holes $L$ diverges to infinity in comparison with their Schwarzschild radius 
\bea
\frac{L}{M} \propto M^{-3/2} \to \infty \, ,
\eea
which implies that direct gravitational interactions between the black holes should become negligible.
\end{enumerate}
The first observation may lead to the conclusion that the dynamics of the background, homogeneous FLRW geometry might decouple from the complicated dynamics of the local inhomogeneities. The second one implies that the matter  on the cosmological scales can be approximated by a collection of a large number of non-interacting massive particles.  

The problem with the reasoning above is that it lacks a solid mathematical basis. The first obvious question is in what mathematical sense we can expect the metric of the BHL to converge to the Einstein-de Sitter (or any FLRW) limit. Obviously the standard pointwise limit does not work, since in the vicinity of each BH the metric strongly diverges from the background. Looking at the problem from the cosmological scale of the background EdS solution we therefore have an ever increasing number of places where the difference between the metrics diverges.

Things become even more complicated if we consider the first and the second derivatives of the four-dimensional metric components, $g_{\mu\nu}$.  The differences between the first derivatives of the two metrics are growing even faster than the differences in the metric itself due to the shrinking of the scale of a single cell, further spoiling the convergence in the vicinity of the black holes. In the second derivatives, we do not see any convergence at all: the underlying EdS model has a non-vanishing Ricci tensor due to the presence of matter, but the BHL is strictly vacuum.

Before going further let us note that the compactness parameter $\mu$, defined in (\ref{eq:mudef}), can serve as a measure of the granularity of the matter distribution.
Namely, the continuum limit  corresponds to $\mu \to 0$ with $\rho = \const$, which implies the scalings $M = \rho^{-1/2}\,\mu^{3/2}$ and $L = \rho^{-1/2}\,\mu^{1/2}$.

\begin{figure}
\centering
  \subfloat[A slice through the 5-cell solution]{\label{5tetrafig}
    \includegraphics[height=2in]{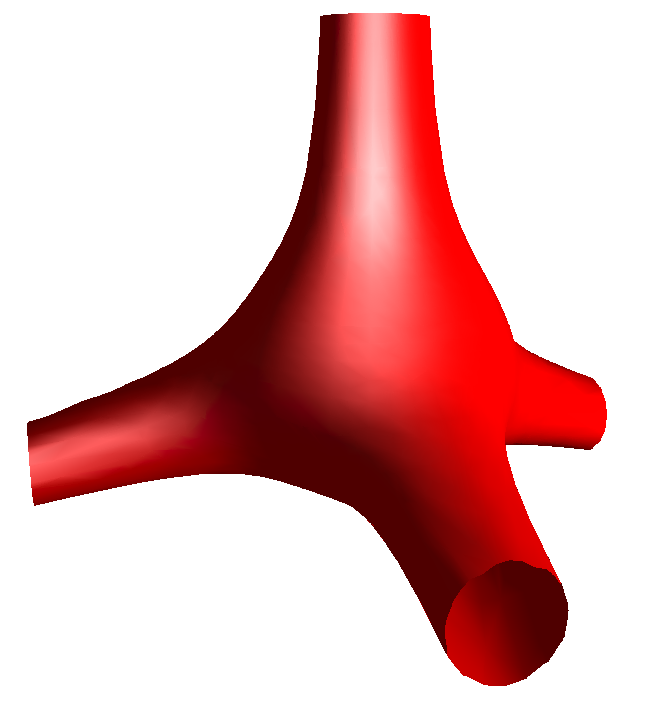}}\qquad\qquad
  \subfloat[A slice through the 8-cell solution]{\label{8cubefig}
    \includegraphics[height=2in]{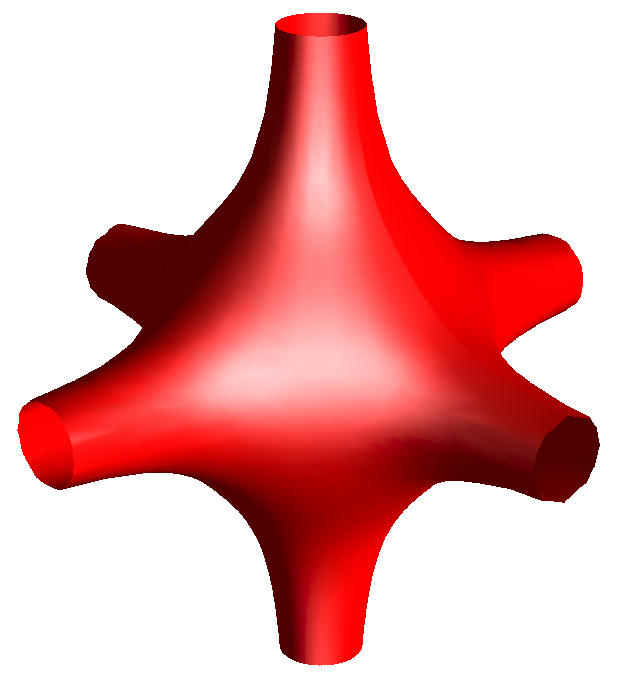}}\\
  \subfloat[A slice through the 16-cell solution]{\label{16cell}
    \includegraphics[height=2in]{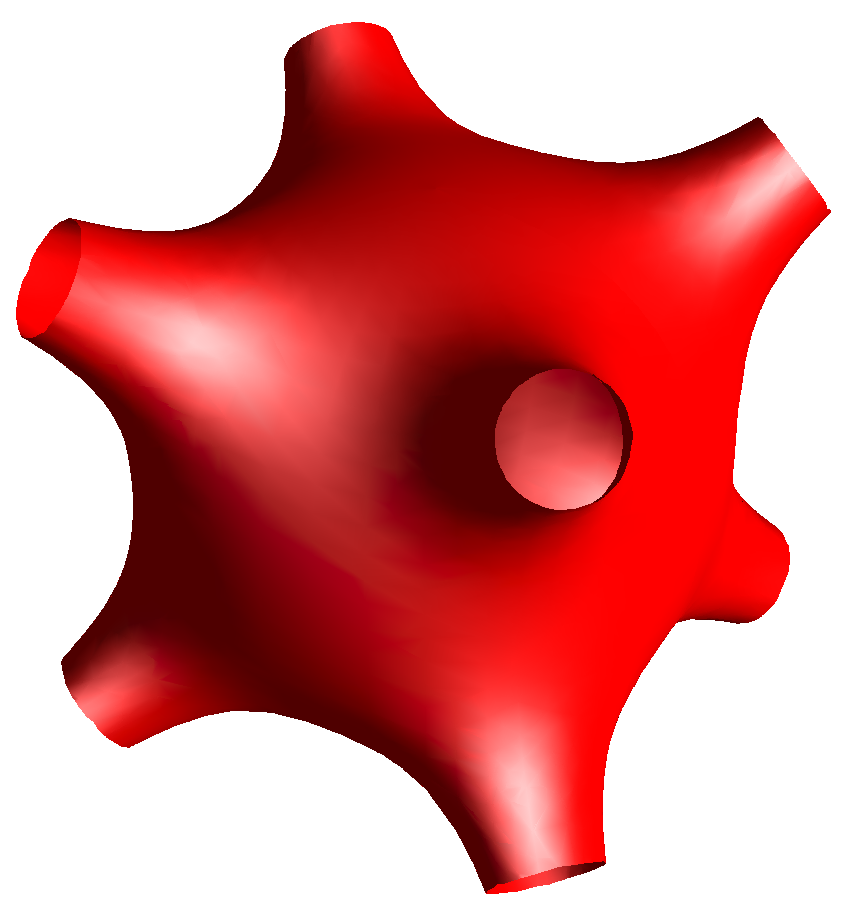}}\qquad\qquad
  \subfloat[A slice through the 24-cell solution]{\label{24cell}
    \includegraphics[height=2in]{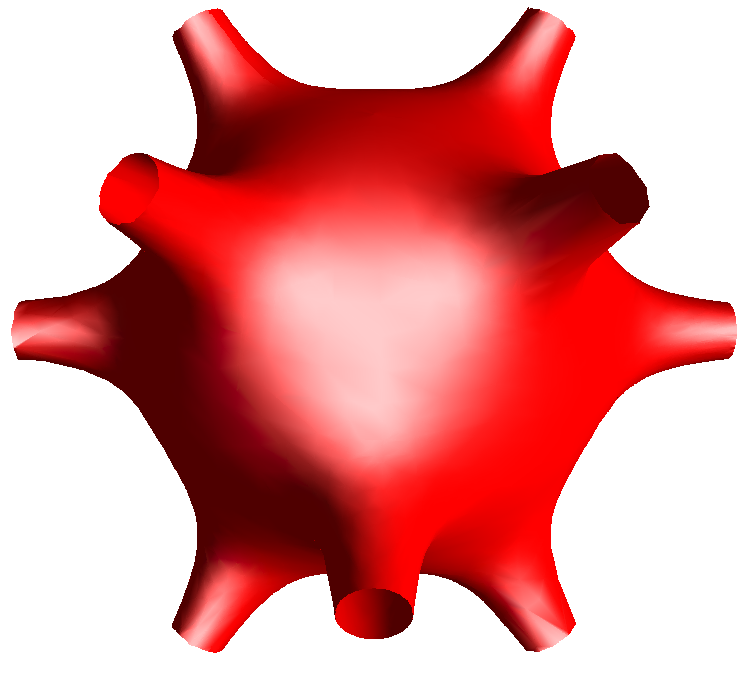}}\\
  \subfloat[A slice through the 120-cell solution]{\label{120cell}
    \includegraphics[height=2in]{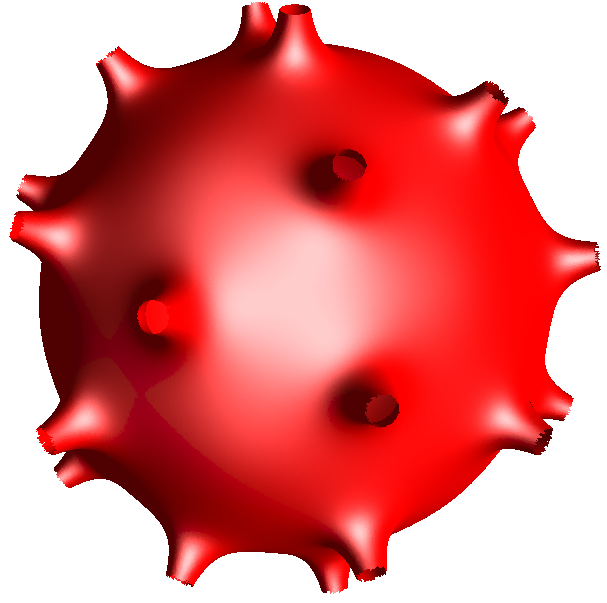}}\qquad\qquad
\subfloat[A slice through the 600-cell solution]{\label{600cell}
    \includegraphics[height=2in]{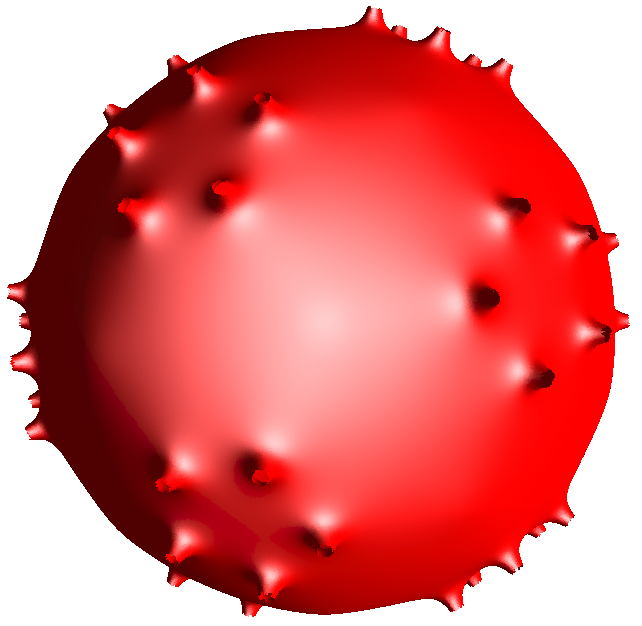}}
  \caption{ A graphical depiction of the scale factor $\psi$, for the six BHL models that correspond to the regular tilings of the round 3-sphere. Each surface corresponds to a slice through the 3-dimensional initial hypersurface, and the value of $\psi$ is given by the distance of the surface from the origin.
  As the number of black holes grows the geometry between the black holes becomes increasingly round, while 
  the spikes, corresponding to the singularities in the conformal factor, become narrower.
Reproduced from \cite{Clifton:2012qh}.
}
\label{fig:6bhls}
\end{figure}

This problem was first discussed in the context of the time-symmetric \Sthree data introduced in Section \ref{sec:momentarily}. Clifton {\it et al} discussed in \cite{Clifton:2012qh} all the configurations corresponding to BHs at the vertices of all 4-dimensional polytopes inscribed in the 3-sphere. The authors noticed that for $N=600$ the geometry of the 3-slice resembles a round sphere with 600 extremely thin spikes, see Figure \ref{fig:6bhls}. On the other hand the deformation of the geometry between the BHs was largest for the smallest $N=5$. Unfortunately it is impossible to probe the limit of $N\to \infty$ this way, since there are only six possible polytopes in four dimensions. 

%

In \cite{Korzynski:2013tea}, Korzy{\'n}ski circumvented this problem by considering more general arrangements of black holes in the time-symmetric, spherical initial data of Section \ref{sec:momentarily}, with
conformal factor given by (\ref{eq:psiclosedform}). The main result of this paper was the derivation of two inequalities. The first bounds from above the deviation of the conformal factor $\psi$ from an appropriately defined global average at a point $p$. The bound depends on the positions and masses of the black holes and the coordinate distance from the nearest black hole, $\lambda_{min}$. 
The dependence on the black hole arrangement is via the modified cap discrepancy, $E$, which measures the difference between the uniform measure on the 3-sphere and a discrete measure concentrated in the singularities of the conformal factor:
\bea
\frac{\psi(p)-\left\langle \psi \right\rangle}{\left\langle \psi \right\rangle} \le C_\varepsilon\,U(E,\lambda_{min}), \label{eq:inequality1}
\eea
with $\varepsilon$ being a small positive number, $C_\varepsilon$ a $\varepsilon$-dependent constant and $U$ a fairly complicated function (see \cite{Korzynski:2013tea} for details).

The inequality above is the key to understanding the geometry of the initial data for a large number of 
black holes. To see why, fix $p$ and consider a sequence of black hole configurations with $N\to\infty$. 
If the black holes are distributed evenly across the 3-sphere then for large $N$ one can prove that $E$ scales like $N^{-1/3}$ or slower. The coordinate distance to 
the nearest black hole on the other hand converges as $N^{-1/3}$ or faster, depending on $p$. It turns out that the convergence of the right-hand side of (\ref{eq:inequality1})
to zero depends on the behaviour of the product $E\left( 1+2E\,\lambda_{min}^{-\frac{1+\varepsilon}{2}}\right)$: if it converges to zero, then so does the deviation of the metric from the FLRW mean. 
This in turn depends on the rate of convergence of $E$ and $\lambda_{min}$ (too slow convergence of $E$, or too rapid convergence of $\lambda_{min}$, will spoil the convergence of the metric).

\begin{figure}[tbp]
    \centering
  \includegraphics[width=0.25\textwidth]{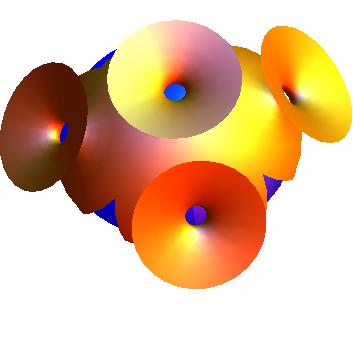}\includegraphics[width=0.25\textwidth]{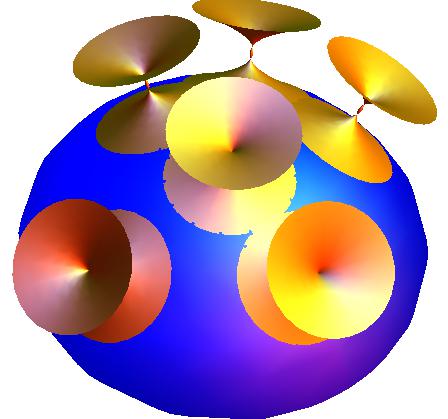}\includegraphics[width=0.25\textwidth]{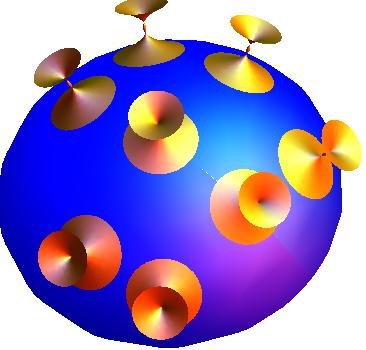}
     \caption{\label{fig:zoomin} A simplified, pictorial representation of the geometry of the initial data on \Sthree with punctures,
     with an increasing number of uniformly distributed black holes: the solution is asymptotically dominated by the faraway region (blue), 
     in which the geometry is approximately spherical, and very close the the FLRW model with a uniform matter distribution. Close to the punctures (yellow) the geometry
     strongly diverges from the background FLRW, and for $N\to\infty$ and sufficiently pairwise separated black holes, it asymptotically approaches that of Schwarzschild.}
\end{figure}

A more refined analysis of concrete examples reveals the following picture: for an even distribution of black holes, and for large $N$, the solution can be divided into 
two regions. The first one is the ``faraway region'', in which the distance to the nearest black hole is much larger than the typical Schwarzschild radius, $R_{\rm S}$, of the black holes. 
We can use inequality (\ref{eq:inequality1}) to show that in this region the metric tensor converges to the metric of a round 3-sphere, i.e. the initial data for the homogeneous 
FLRW solution. In other words, for $N\to\infty$ the geometry in this region seems not to notice the discreteness of the matter distribution. On the other hand, in the region 
located within the distance of a few Schwarzschild radii from the nearest black hole the geometry is dominated by its presence and deviates strongly from the background FLRW
metric for all $N$. If the black hole in question is isolated, i.e.~located very far from the nearest neighbour in terms of $R_{\rm S}$, the metric converges to the Schwarzschild solution
in its vicinity. Ultimately, it can be shown that, as $N$ diverges, the first region asymptotically dominates the whole solution in terms of volume (see Figure~\ref{fig:zoomin}). In other words, and somewhat
surprisingly, despite the number of black holes diverging to infinity, almost all of the solution turns out to lie very far away from the nearest black hole. 

Thus the continuum limit of the metric arises as the limit attained by the metric only in the empty space between black holes. However, the set of points in which this convergence
works 
grows with $N$ and asymptotically takes over the whole solution save for a countable number of points. Note that this picture is entirely consistent with the
observations made for regular lattices on \Sthree in \cite{Clifton:2012qh}.
 
A similar inequality was also proven for the deviation of the sum of the black hole masses, $M_{\tot}$, from the effective mass, $M_{\eeff}$, inferred from the large-scale geometry: 
\bea
 \frac{\left|M_{\eeff} - M_{\tot}\right|}{M_{\eeff}} \le  C_\varepsilon\,W_\varepsilon\left(E,\frac{\alpha_{\max}}{\alpha},\delta_{\max},\delta_{\min}\right),
 \label{eq:inequality2}
\eea
where $E$ is again the discrepancy, where $\delta_{\max}$ and $\delta_{\min}$ are the largest and the smallest distances between a pair of black holes, and where $\alpha_{\max}$ is the
largest $\alpha_i$ parameter of a single black hole (as introduced in equation (\ref{eq:psiclosedform})), while $\alpha = \sum_{i=1}^{N} \alpha_i$.  If the left-hand side converges to zero in the continuum limit, then asymptotically
the total mass is simply the sum of masses of the individual black holes. Thus all backreaction effects vanish in the $N \to \infty$ limit.

In the continuum limit the right-hand side of this inequality is 
then bounded from above by the product $E\left(1 + E\,\delta_{\min}^{-\frac{1+\varepsilon}{2}}\right)^2$. Once again the convergence of the left-hand side depends
on the rates of convergence of $E$ and $\delta_{\max}$ to zero. If we make sure that the black holes are spaced far enough from each other, then the convergence of the 
minimal distance between black holes will be slow in comparison with $E$. Thus, the inequality (\ref{eq:inequality2}) can be used to prove that within this class of models
the backreaction vanishes in the continuum limit, if we make sure that the black holes on the finest scales do not cluster, but rather stay asymptotically far from each other. The case of clustering black holes in considered further in Section \ref{sec:struct}.

\subsection{Approximation schemes}
\label{sec:approx}

The field of BHLs studied by the means of approximation schemes was pioneered by Lindquist and Wheeler \cite{RevModPhys.29.432}. Their approach was based on a ``rough'' matching of 
spherically symmetric cells, and therefore lacked a solid geometric background.  It was revisited by Clifton and Ferreira \cite{Clifton:2009jw,Clifton:2010fr}, Clifton, Ferreira and O'Donnell \cite{Clifton:2011mt}, and Liu~\cite{Liu:2015bya}, who studied the dynamics and the optical properties of those approximate models. 
A more mathematically sound approach, based on rigorous perturbative treatment (expansion in $\mu^{1/2}$) and with cubic lattice geometry, was
presented by Bruneton and Larena \cite{Bruneton:2012cg, Bruneton:2012ru}. The authors managed to derive the leading-order behaviour of the dynamics, consistent with an 
appropriately fitted Einstein-de Sitter solution, and studied the optical properties of cubic lattices.

Sanghai and Clifton proposed in \cite{Clifton:2010fr}, \cite{Sanghai:2015wia} and \cite{Sanghai:2016ucv} a new method of constructing 
approximate, inhomogeneous solutions of Einstein's equations with a discrete symmetry. Their method is based on patching together identical cells of sub-horizon size. The authors assume for simplicity that the cell faces are reflection symmetric, and that within the interior of each cell Einstein's equations can be solved using the post-Newtonian approximation scheme. As these authors note, inhomogeneous models are usually constructed in a top-down way, by  introducing inhomogeneities in a given background FLRW solution. This can be done in either the standard perturbative approach, or by gluing in another solution (as in the Swiss-cheese models). Sanghai and Clifton 
turn this approach around and construct their model bottom-up, i.e.~they begin by modelling a single cell of a periodic lattice. The cell serves as a building block for a spacetime composed
of identical elements. It turns out that the consistency conditions imposed on the geometry of the boundary then determines the large-scale dynamics of the lattice. It is found that
the lattice geometry must follow the standard FLRW dynamics with dust at the lowest order in the post-Newtonian expansion. Thus the ``background FLRW'' geometry  is not imposed {\it a priori}, but rather arises in a natural way as an ``emergent cosmology'' when we consider the dynamics of the cell faces.

The cells in this approach can take the shape of any convex polyhedron capable of tiling a constant curvature 3-space. The Israel matching conditions for reflection-symmetric surfaces (the vanishing of the extrinsic curvature and the continuity of the induced 3-metric) are then assumed to hold on the faces of each of these cells, and the area of a single cell face can be used as a measure of the cell size (as mentioned in Section \ref{sec:evol}). Einstein's equations are solved inside each cell perturbatively, with the Minkowski metric serving as the background. 
Unlike the standard cosmological perturbation theory, this approach does not require the density contrast $\delta$ to be small, and the formalism can work even if the
whole matter content of the lattice cell is squeezed into a small structure at the centre.

The authors focus on the geometrically simplest case, i.e. on a flat cubic lattice. As usual in the post-Newtonian formalism, the perturbative expansion is governed by the parameter $\epsilon = v/c \ll 1$, where $v$ is the velocity of matter. Einstein's equations in this case reduce to the Poisson equations for each of the metric components, with the matter, momentum and stress density as sources, all solved with periodic boundary conditions. It turns out that in the leading, Newtonian order the size of the cell size must follow the evolution of a flat homogeneous model (Einstein-de Sitter) with the effective energy density of the model is given by the standard integral of the matter density. The matter velocity field plays no role at this order.

The whole system of Einstein's equations plus matching conditions may then be solved perturbatively in the next orders in a consistent manner. The authors obtain higher-order corrections for the effective field equations, i.e.~the backreaction terms, as integrals of the first-order potentials. They take the form of terms modifying the effective gravitational mass in the effective Friedman equations, as well as terms that take the form of an effective radiation fluid. These corrections are evaluated precisely in the case of a lattice composed of point sources, i.e. matter arranged in the form of a three-dimensional Dirac comb:
\bea
 \rho(x^i) = \sum_{\beta \in \ZZ^3} M\,\delta(x^i - 2L\beta^i).
\eea
In the linearised theory this distribution corresponds to a black-hole lattice, although 
in the full theory this type of singular matter distribution is divergent. Fortunately, at the level of the post-Newtonian expansion required to obtain the leading-order backreaction effects, 
the nonlinear terms are finite.

The reader may note that this formalism crucially requires that the size of the cell is much smaller than the Hubble scale of the large-scale FLRW solutions. Therefore the solutions constructed in this way correspond to the $\mu \ll 1$ limit discussed in the section above. Failing to obey this requirement results in the loss of consistency of the approximation on the cell boundary. 

The results of this construction are intriguing: in~\cite{Sanghai:2015wia}, the size of the boundary follows the Friedman equations with dust and a small
backreaction term in the form of a fictitious radiation with negative energy density. The backreaction term arises from the nonlinearity of Einstein's equations and is proportional to $\mu^2$.  
In \cite{Sanghai:2016ucv}, this result is generalised by considering the bottom-up construction of a BHL with spatial curvature, a cosmological constant and a radiation fluid. Again the backreaction term is evaluated and it turns out that it again mimics additional radiation at leading-order. 
The cosmological constant is found to have a very small impact on the backreaction term, while the presence of radiation tends to decrease backreaction in both the first and the second Friedman equation. The spatial curvature can have either a damping or an amplifying impact on the backreaction, depending on its sign.

\section{Effective properties of a BHL}
\label{sec:effect}

A central feature of a discrete cosmology, in addition to its kinematical properties, are its effective physical properties (such as density and pressure) on large scales. In this section we will outline how global concepts, such as the total mass in the universe, are not necessarily given by simply averaging or summing the masses of every body. Instead, we suggest that dressed parameters should be considered, which can contain information not only about the mass of individual black holes, but also about their interactions with the other bodies that exist in the universe.

\subsection{Coarse graining and dressing}
Coarse graining processes are commonly used in many areas of physical modelling where effective theories are used, and in this section we will borrow some of the frequently used terminology. In particular, we will be interested in the difference between \emph{bare} and \emph{dressed} quantities. Of course, the existence of different definitions of mass is well-established in General Relativity, and can be observed even in relatively simple systems, such as isolated binaries. In these systems the total mass (as measured at infinity) is different from the sum of the individual local masses that one would infer from the geometry of space-time near each of the bodies individually (using, for example, the Isolated Horizon formalism). In a cosmology constructed from multiple black holes it is interesting to ask which measure of mass we should expect to occur in the effective Friedmann equation, which describes expansion on the largest scales.

As in the case of isolated binaries, there are multiple possible definitions of mass that we could use, and we need to be able to identify which of them it is sensible to use in any given circumstance. Further, we may also want to understand what the physical and mathematical differences are between difference concepts of mass in BHL cosmologies. For example, we mentioned in the sections above that an FLRW counterpart to a BHL in the \Sthree class can be defined as the one where curves of identical coordinate specification have the same proper length at a chosen time. The mass difference the BHL and the FLRW model can then be interpreted as the effect of backreaction (as was done in~\cite{Bentivegna:2012ei}). Consistent with the findings in~\cite{Clifton:2012qh}, one finds that a closed FLRW universe of the same scale contains more mass than the corresponding BHL. Inhomogeneities, in this case, appear to act to increase the effective mass of the system with respect to its homogeneous counterpart. This is illustrated in the first column of Table~\ref{tab:s3mass}.

\bt
\center
\begin{tabular}{|c|c|c|}
\hline
 \phantom{{\LARGE A}} Number of black holes \phantom{{\LARGE A}} & $ \frac{M_{\rm ADM}}{M_{\rm global}}$ & $\frac{M_{\rm ADM}}{M_{\rm effective}}$ \\[5pt]
\hline
   5  & 0.74 & 5.0         \\
   8  & 0.80 & 9.1         \\ 
  16  & 0.91 & 22         \\
  24  & 0.91 & 34         \\
 120  & 0.97 & 190        \\
 600  & 1.00 & 1000         \\
\hline
\end{tabular}
\caption{ADM, global and effective masses of the six regular \Sthree BHLs. The {\it ADM mass} represents the sum of the masses of each individual black hole, each measured as if they were in isolation (i.e. on the corresponding asymptotically-flat end of the Einstein-Rosen bridge). The {\it global mass}, on the other hand, is the mass of a closed FLRW model with the same initial scale, and the {\it effective mass} is the mass including all interaction energies. The discrepancy between these measures represents a dressing effect due to the discrete nature of the mass distribution (see~\cite{Clifton:2012qh,Bentivegna:2012ei,Clifton:2014mza} for details). \label{tab:s3mass}}
\et

%

The same process can be performed with the \Tthree class of BHLs, where one is again at liberty to ask what are the physical properties of the fitted FLRW models. This question is investigated in detail in~\cite{Bentivegna:2013jta}, where effective fluid quantities are defined based on the evolution of the edge length of the cell. In this case one finds that the mass of a cube in the corresponding FLRW cosmology has to be smaller than the mass of the central black hole in discrete cosmology (an effect that has opposite sign to the one in \Sthree). One should note, however, that the black hole mass estimate is based on the horizon mass for the \Tthree case, as the ADM mass is not available. Other quantities, such as the pressure in the best-fitting FLRW model, remain quite close to zero.

Going further, one can identify the ADM mass of each black hole with the ``proper mass'' of the body (something which has been done explicitly in some parts of the literature, e.g. \cite{Clifton:2012qh}). The idea behind this is to have a measure of the mass that the black hole would have if it was considered in isolation. This is in contrast to the ``effective mass'', which is the mass that an observer would infer if the body were to be placed in a gravitationally interacting system with other massive bodies. The effective mass should then be expected to contain contributions not only from the proper mass of the body, but also from the gravitational interactions between the body in question and any other bodies that may also exist. In this context, the effective mass is the dressed parameter, and the proper mass is the bare parameter.

Determining interaction energies in gravitational physics is most easily done at infinity. In the case of a BHL, however, there is no region that is infinitely far from all masses. This causes some ambiguity in the definition of the effective mass -this is discouraging, but should not nullify the physics behind the idea. A first attempt at identifying the effective mass in a BHL was performed in \cite{Clifton:2012qh}, where the authors proceeded in analogy to the mathematical procedure that occurs when infinity is present. The results were somewhat surprising, as effective mass was found to be very small compared to the proper ADM mass, and became smaller as the number of bodies in the universe was increased. See the last column of Table~\ref{tab:s3mass}, for the case of the six regular BHLs in $S^3$. This result can be understood by recalling that gravitational potential energies have a negative sign (while mass is always positive), and by noting that the number of pairwise interactions between masses increases rapidly as the number of bodies in the universe increases.

The idea of gravitational interaction energies is cosmologies with discrete bodies was fleshed out a little further in \cite{Clifton:2014mza}, where the method of images was used to create clusters of masses, as well as multiple cosmologies connected together by throats. The interaction energies within a given cosmology could then be inferred by using the observers in the other cosmological regions. These ideas were built on further in \cite{Durk:2017rky}, where interactions both within and between clusters of masses were considered, as discussed below. The results of these studies appear to confirm that the sum total of all interaction energies in a BHL cosmology does indeed appear to become large as the number of bodies increases, but that not all interaction energies contribute to the effective Friedmann equations.

\subsection{Structuration and interaction energies}
\label{sec:struct}

The original six lattice configurations on positively closed backgrounds, discussed in Section \ref{sec:arr}, can also be extended to include the consequences of the existence of large-scale structures. Such investigations allow the subtleties of mass dressing to studied in more detail, as the formation of structure alters the interaction energies between masses, and allows the dressed mass parameters to change accordingly.

The formation of structure in momentarily static initial data has been investigated in \cite{Durk:2017rky}, where structuration was included in a very specific way: Each black hole positioned at the centre of each primitive lattice cell was split into the corresponding number of vertices of that cell, and each of these new masses was then moved along an LRS trajectories towards one of those vertices. The paths that the masses followed were parameterized by $\lambda \in [0,1]$, such that for values of $\lambda \sim 0$ the masses have moved only a small distance from their original position, and for $\lambda \sim 1$ they approached the dual lattice (as given in Table \ref{tab:numbers}). At intermediate values of $\lambda$ the masses are far apart, and the configuration represents a cluster of black holes around the position of either the original lattice or its dual. In this way the parameter $\lambda$ controls the degree of clustering in these models, and creates a family of models that approaches the original lattice configurations as $\lambda \rightarrow 0$ or $1$.

\begin{table}[tbp]
\centering
\begin{tabular}{|c|c|c|c|c|}
\hline
Original $N^{\underline{o}}$ & Cell shape & $N^{\underline{o}}$ vertices & Intermediate $N^{\underline{o}}$  & Dual $N^{\underline{o}}$ \\
 \hline
5 & Tetrahedron & 4 & 20 & 5\\
8 & Cube & 8 & 64 & 16\\
16 & Tetrahedron & 4 & 64 & 8\\
24 & Octahedron & 6 & 144 & 24\\
120 & Dodecahedron & 20 & 2400 & 600\\
600 & Tetrahedron & 4 & 2400 & 120\\
\hline
\end{tabular}
\caption{\label{tab:numbers}The numbers of masses for each of the new structured models (original, intermediate and dual), and the number of vertices and shape of each of the original lattice cells.}
\end{table}

\begin{figure}[tbp]
    \centering
  \includegraphics[width=0.8\textwidth]{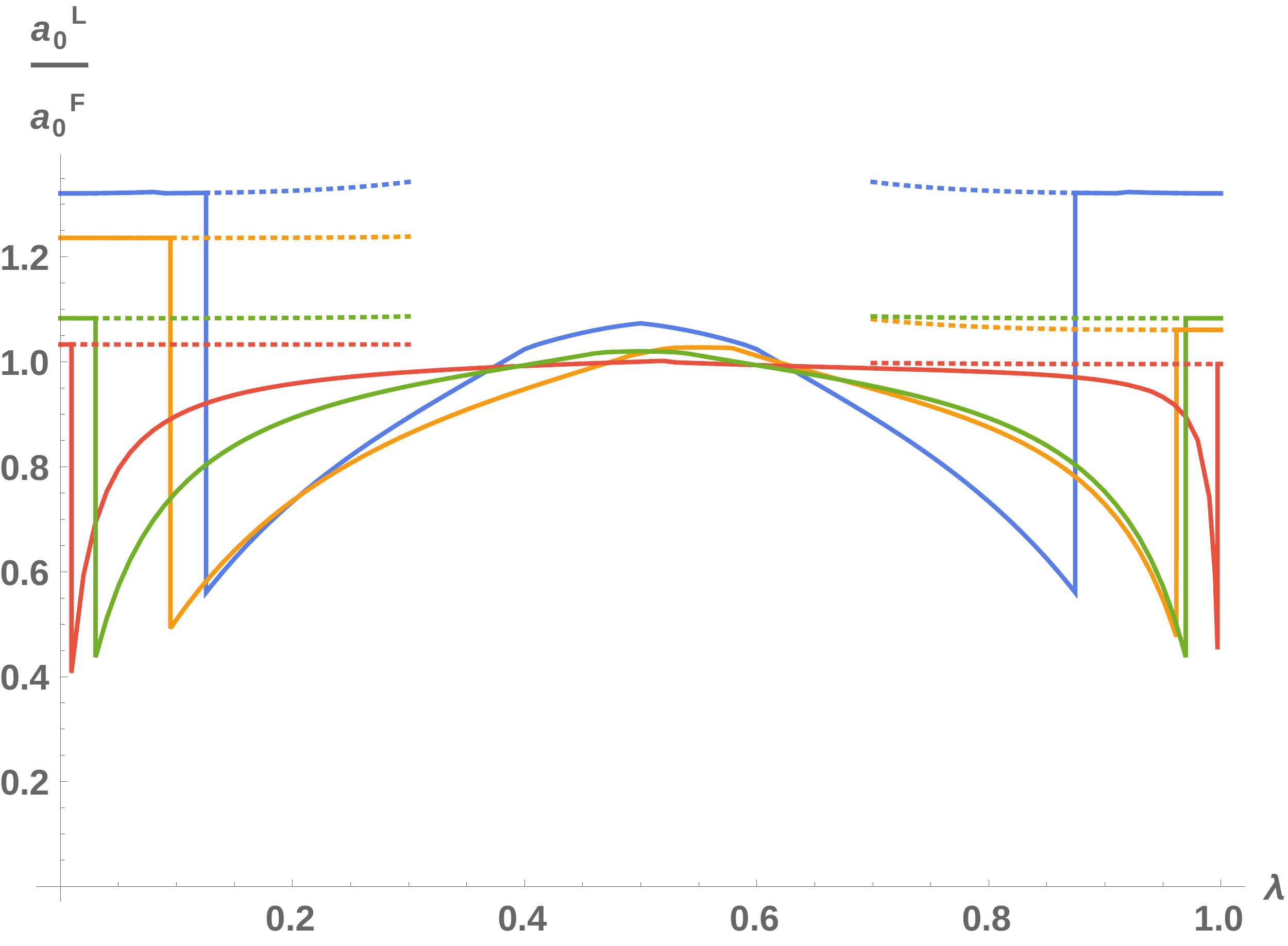}
     \caption{\label{fig:scales2}The ratio of scale factors between the lattice universes, $a_0^L$, and the corresponding dust-dominated positively-curved FLRW universes, $a_0^F$, as a function of $\lambda$. The four curves correspond to the models with the following number of masses: $5 \to 5$ (blue), $8 \to 16$ (orange), $24 \to 24$ (green), and $120 \to 600$ (red). The black holes are separated by the largest distance when $\lambda \sim 0.5$, and are tightly clustered when $\lambda \sim 0$ or $1$. Reproduced from \cite{Durk:2017rky}.}
\end{figure}

In order to perform comparison between these clustered models, and their corresponding FLRW counterparts, we more or less proceed as in section \ref{sec:evol}, and consider closed FLRW universes with the same total proper mass. In order to find an analogue of the scale factor in the black hole cosmology we look for the value of $\psi$ at the location where it achieves its global minimum, which in some sense corresponds to the point furthest away form all of the black holes (unlike the cell edge prescription, this can be achieved for any given value of $\lambda$). The resulting ratio of scale factors is shown graphically, as a function of $\lambda$ in Figure \ref{fig:scales2}. 

The sudden drops that appear in this figure are at points where a shared horizon forms around the clustered masses. When this happens individual masses are no longer identifiable as separate objects for observers in the cosmological region, and the mass of the cluster must be inferred from their joint gravitational field. The dotted lines correspond to the results one would obtain by taking into account not only the sum all individual black hole proper masses, but also the gravitational effects of their interaction energies with other masses within the cluster.

It can be seen that if interaction energies are ignored, and the black hole universe is compared to an FLRW model with the same total proper mass only, then discretising the matter content has the largest effect when the black holes are separated enough to be considered as individual black holes, but are still very close together (i.e. after the sudden drops that are visible in Figure \ref{fig:scales2}). At this point the lattice universes can be less that 50\% of the size of their FLRW counterparts. On other hand, taking into account the intra-cluster interaction energies between nearby masses when calculating the total mass in the cosmology gives quite different results. In this case the effect of allowing structuration to increase has a much less marked effect, as is clear from the dotted lines in Figure \ref{fig:scales2}.

This study appears to show that neglecting interaction energies, when comparing cosmological models, can cause very large discrepancies in the global scale when comparing lattice models and dust-filled FLRW models. Instead of using ADM masses, it therefore appears that one should use a dressed measure of mass, that contains some contribution from the interaction energies with other masses in the universe. While it is not surprising that gravitational interaction energies can gravitate in general relativity, it is interesting that these appear to be needed in order to find a sensible best-fitting Friedmann cosmological model. This is because gravitation interaction energies are routinely neglected in perfect fluid dust cosmologies, and because the sum of all interaction energies in closed lattice cosmologies appears to be very large indeed \cite{Clifton:2012qh}. 

Similar conclusions were reached by Korzy\'{n}ski in \cite{Korzynski:2014nna}, where series of nested structures were included in cosmological models by placing constant density ``caps'' over the central region of Schwarzschild masses. These caps were then used to embed further layers of structure, and the process was continued over and over again, leading to structures nested within structures. If sufficiently many layers of this type of structuration were included, it was found that the global cosmology could be shown to be quite different to an FLRW model with the same total ADM mass. The reason for this is ultimately the same as above: The interaction energies between the masses were themselves gravitating, and contributing to the total energy content of the space-time. This further supports the idea that dressed mass parameters should be used in cosmology when structure is present, in order to determine the total energy budget of the universe.

\section{Optical properties}
\label{sec:opt}

The optical properties of a spacetime are of critical importance for observations, as almost all astronomical inferences about cosmology are based on the collection of light. In this regard, the most basic of the quantities that astronomers measure are redshifts, luminosity distances and angular diameter distances of astrophysical sources such as galaxies and supernovae. It is the interpretation of these quantities, in the form of Hubble diagrams and other more complicated constructs, that allows us to infer the large-scale properties of the Universe and the matter that fills it. In this section we will review some of the old arguments involving the effects of inhomogeneity on the optical properties of a spacetime, before moving on to summarise new results involving redshifts and distance measures within lattice cosmologies specifically.

\subsection{Ray optics}

The fundamental equations that govern the propagation of light within general relativity are the Einstein-Maxwell system. However, the solutions to the full equations can be somewhat complicated, due to the non-linearities of Einstein's equations, and the fact that the electromagnetic fields themselves are a source of gravitation. It is therefore common to use the {\it geometric optics} approximation when considering the optical properties of cosmological models. This approximation considers the limit where (i) the wavelength of the electromagnetic waves is short compared to the curvature scale of the spacetime, and (ii) the amplitude of the waves can be considered small in some suitably defined sense. If these conditions are met, then the gravitational fields of the electromagnetic waves can be neglected, and the waves themselves can be seen to follow null geodesics.

Let us now consider a set of light rays, that obey the approximations described above, and that have a tangent vector field $k^a$. The conditions that the rays of light are geodesic and null immediately imply
\beq
k^a k^b_{\phantom{b} ; a} =0 \qquad {\rm and} \qquad k^a k_a =0 \, .
\eeq
The observed frequency of the light that follows the integral curves of this vector field depends on the normalised 4-velocity of the observer, $u^a$, and be written as $ -u^a k_a$. This immediately allows us to define a redshift in the photon's energy, between emission and observation, as follows:
\beq
1+z = \frac{(-u^a k_a)\vert_e}{(-u^b k_b)\vert_o} \label{z} \, ,
\eeq
where subscripts $e$ and $o$ denote that a quantity is determined at the emission and observation events, respectively. As soon as the world-lines of the source and observer are chosen, the equations above can be used to determine the null geodesic curves that connect them, and the redshifting of the photons between emission and observation.

The calculation of measures of distance requires us to know not only the properties of individual rays of light, as is sufficient to find the redshift, but also the rate at which neighbouring light rays diverge away from each other. This behaviour is essentially encoded entirely within the geodesic deviation equation, which for rays of light separated by an infinitesimal distance reads
\beq
\frac{D^2 \xi^a}{d\lambda^2} = R^{a}_{\phantom{a} b c d} k^b k^c \xi^d \, ,
\eeq
where $\xi^a$ is a vector that connects the neighbouring geodesics and $D/d\lambda = k^a \nabla_a$ is a derivative with respect to the affine distance $\lambda$ along the rays. To tease out the different deformations that are possible for a given beam of light it is convenient at this point to introduce a set of orthonormal basis vectors: $\{u^a, d^a, s_1^a, s_2^a\}$, where $k^{a} = (-u^b k_b) (u^a + d^a)$. The space-like unit vector $d^a$ then picks out the direction of propagation of the photon in the frame of the observers that follow $u^a$, and the remaining vectors $s_1^a$ and $s_2^a$ span the space of 2-dimensional screens that are orthogonal to this direction. This basis can be used to define a projection tensor, $S_{ab} = g_{ab} +u_{a} u_{b}- d_{a} d_{b}$, which projects onto the screen space spanned by the real and imaginary parts of the complex basis vectors $m^a=\frac{1}{\sqrt{2}} (s_1^a + i s_2^a)$. By construction, these vectors obey $m^a u_a=0$, $m^a d_a=0$, $m^a m_a =0$ and $m^a \bar{m}_a=1$, and can be chosen to be propagated along the null rays according to $S^{a}_{\phantom{a} b} k^c m^{b}_{\phantom{b};c} = 0$.

This machinery is very useful as it allows us to classify the deformation of the beams of light in terms of the morphology of the shadow they leave on the screens that are carried by observers. For these purposes, it is convenient to define the following two real scalars: $\theta = \Re ( k_{a ; b} m^a \bar{m}^b)$ and $\omega = \Im ( k_{a ; b} m^a \bar{m}^b)$, which describe the extent to which the beam of light expands and rotates as it propagates through space. A third complex quantity, $\sigma = -k_{a;b} m^a m^b$, describes the area-preserving rate of shear of the beam. Together, these three quantities provide a full description of all of the possible behaviours of a set of light rays that are separated by infinitesimal distances. However, if the rays of light are initial non-rotating, because they are emitted from a point mass (for example), then we can set $\omega=0$. The propagation equation for $\omega$ is such that if it starts off non-rotating, then it stays non-rotating forever. For this reason the rotation of beams of light are usually neglected in cosmology, which leaves us with only the expansion and the shear.

The evolution equations for the expansion and shear scalars, in the absence of rotation, are determined from the geodesic deviation equation, and are given as follows:
\bea
\frac{D\theta}{d\lambda} + \theta^2 +\bar{\sigma} \sigma = -\frac{1}{2} R_{ab} k^a k^b \label{dth}\\ 
\frac{D\sigma}{d\lambda}+2 \sigma \theta = C_{abcd} k^a m^b k^c m^d \, , \label{dsig}
\eea
where overbars denote complex conjugation, and $C_{abcd}$ is the Weyl tensor. Once these equations have been solved, the angular diameter distance and luminosity distance are then given respectively by
\beq
D_A \propto {\rm exp} \left\{ \int_e^o \theta d \lambda \right\} \qquad {\rm and} \qquad D_L = (1+z)^2 D_A \label{ra} \, .
\eeq
These equations are valid in any spacetime, as long as the geometric optics approximation is valid, and can be directly integrated to obtain the distance measures that observers following $u^a$ should be expected to obtain.

\subsection{Empty beams and approximation schemes}

The problem of calculating observables in an inhomogeneous universe has been discussed many times in the literature, but one of the earliest and clearest of these was that presented by Zel'dovich in 1964 \cite{1964SvA.....8...13Z}. This discussion notes that in a perfectly homogeneous and isotropic space the Weyl curvature tensor must vanish completely, while the Ricci curvature can be non-zero. This means that in such spaces the driving term on the right-hand side of the shear evolution equation (\ref{dsig}) must vanish, while the driving term on the right-hand side of the evolution equation for the expansion (\ref{dth}) can be non-zero. A circular beam of light therefore remains circular, as one would expect in spacetime in which there is no preferred direction in space at any point. 

However, in a universe where all matter is clumped into dense objects, such as stars and galaxies, the situation is exactly reversed: The right-hand side of equation (\ref{dth}) must vanish in the regions of vacuum through which the light travels, while the right-hand side of equation (\ref{dsig}) should be expected to be non-zero in general. This means that the beam of light gets distorted as it passes by the massive objects, as the shear will become non-zero even if it starts out at zero. It is then the magnitude of the shear scalar that sources the evolution of the expansion scalar in equation (\ref{dth}), and not the Ricci curvature of the spacetime (which in this case vanishes exactly, if the region outside the massive objects is a perfect vacuum). Equations (\ref{dth}) and (\ref{dsig}) therefore seem to open up the possibility that distance measures could be quite different in spacetimes where the matter is clumped compared to those where it is smoothly distributed, even if the large-scale expansion is identical in both cases. The lattice cosmologies discussed in this review are ideal for studying the possible effects that could arise due to extreme inhomogeneity.

Of course, the results for the real Universe should be expected to lie somewhere between the results from lattice cosmologies and the standard results from Friedmann cosmology, as the real Universe contains both opaque objects (e.g. stars and galaxies) as well as diffuse matter (e.g. gas and dark matter). We should therefore expect both the accumulation of shear and non-zero Ricci curvature, both of which source the evolutions of $\theta$ and therefore cause the focusing of beams of light. In order to model this complex situation a number of approximation schemes and results have been constructed and proven. We will outline some of the most prominent of these here, as they are useful for interpreting the results from numerical integration of the null geodesic equations in lattice spacetimes, and because lattice spacetimes allow for their validity and domains of applicability to be directly investigated.

Probably the most well known approximation scheme for the optics of cosmological spacetime is the Dyer-Roeder approach \cite{1973ApJ...180L..31D}. In this scheme one first assumes that the redshift along null rays is given as a ratio of the cosmological scale factors at the time of observation and emission, just as in Friedmann cosmology. Next, the shear is neglected entirely by assuming that $\sigma=0$ along every ray of light. Lastly, the amount of matter along the light rays is assumed to be some fraction, $\alpha$, of the cosmological average. Under these three assumptions equations (\ref{z}), (\ref{dth}) and (\ref{ra}) can be used to derive
\beq
\frac{d^2 D_A}{dz^2} +\left( \frac{1}{H} \frac{dH}{dz} + \frac{2}{1+z} \right) \frac{d D_A}{dz}  + \frac{3 \alpha \Omega_m H^2_0}{2 H^2} (1+z) D_A =0 \, ,
\eeq
where $0 \leqslant \alpha \leqslant 1$ and where we have assumed the universe to be filled with pressureless matter and a cosmological constant, so the Hubble rate in the corresponding Friedmann universe is given as a function of redshift by $H (z) =H_0 \sqrt{\Omega_m (1+z)^3 + \Omega_k (1+z)^2 + \Omega_{\Lambda}}$. 

This approximation scheme reduces the problem of calculating the angular diameter distance as a function of redshift to solving one simple ordinary differential equation. The constant $\alpha$ gives the assumed amount of matter along the light ray in question. It equals $1$ for a perfect Robertson-Walker geometry, and should be set equal to zero if the beam of light is taken to pass through perfect vacuum. This latter case is referred to as the {\it Empty Beam Approximation}, and gives significantly different Hubble diagrams than distance measures calculated in the standard Friedmann cosmology \cite{1972ApJ...174L.115D}. It does, however, also neglect all of the effects of shear (which should generally be expected to add to the focusing of the beam), as well as any deviations in the redshift relation from the predictions of Friedmann cosmology. In general, $\alpha$ is taken to be a function of $z$ in order to model a time-dependent energy density along the line of sight.

A more sophisticated statement about the averages of the angular diameter distance in inhomogeneous cosmologies has been given by the Weinberg-Kibble-Lieu theorem \cite{1976ApJ...208L...1W,2005ApJ...632..718K}. This goes as follows; The directional average of the square of the angular diameter distance, on surfaces of constant $\lambda$, is unaffected by the presence of inhomogeneities. This statement is valid for weak gravitational fields around Friedmann backgrounds, and is very useful for fully understanding the effects of inhomogeneity on the average of observables that astronomers might measure. In particular, if the redshifts of distant astrophysical objects as functions of the affine distance along the null rays is unperturbed from the Friedmann values, then it gives a measure based on direct observables that is expected to faithfully represent properties of the background cosmology.

Finally, a recent addition to our understanding of the propagation of light in inhomogeneous cosmologies has been made by modeling lensing as a stochastic process \cite{2015JCAP...11..022F}. This Stochastic Lensing Formalism treats the geometric sources of the focusing and deformation of a beam of light as being white noise, and calculates not only the expected mean of the distribution of angular diameter distances along many lines of sight, but also the moments of the distribution around the mean. It is therefore potentially a very useful formalism for understanding observations within an inhomogeneous universe, and could be used to understand and interpret observations within black-hole lattices. 

We will return to all of these approaches below, when we describe the results of directly calculating observables in these spacetimes using both numerical and perturbative methods.

\subsection{Numerical spacetime}

The optical properties of the numerical spacetimes discussed in Section \ref{sec:evol} have recently been studied in \cite{2017JCAP...03..014B}. The authors of this study used code generated within the Einstein Toolkit, based on the \texttt{Cactus} framework with the modules \texttt{Carpet}, \texttt{McLachlan} and \texttt{CT\_MultiLevel}, in order to create numerical models of black-hole lattice cosmologies. They then created a new ray tracing module, which was capable of following the paths of photons within these spacetimes. This additional code was verified against the known results from exact solutions, and then used to explore the optical properties of some geometrically special trajectories within their numerical models.

\begin{figure}[t]
\begin{centering}
\includegraphics[width=0.5\textwidth]{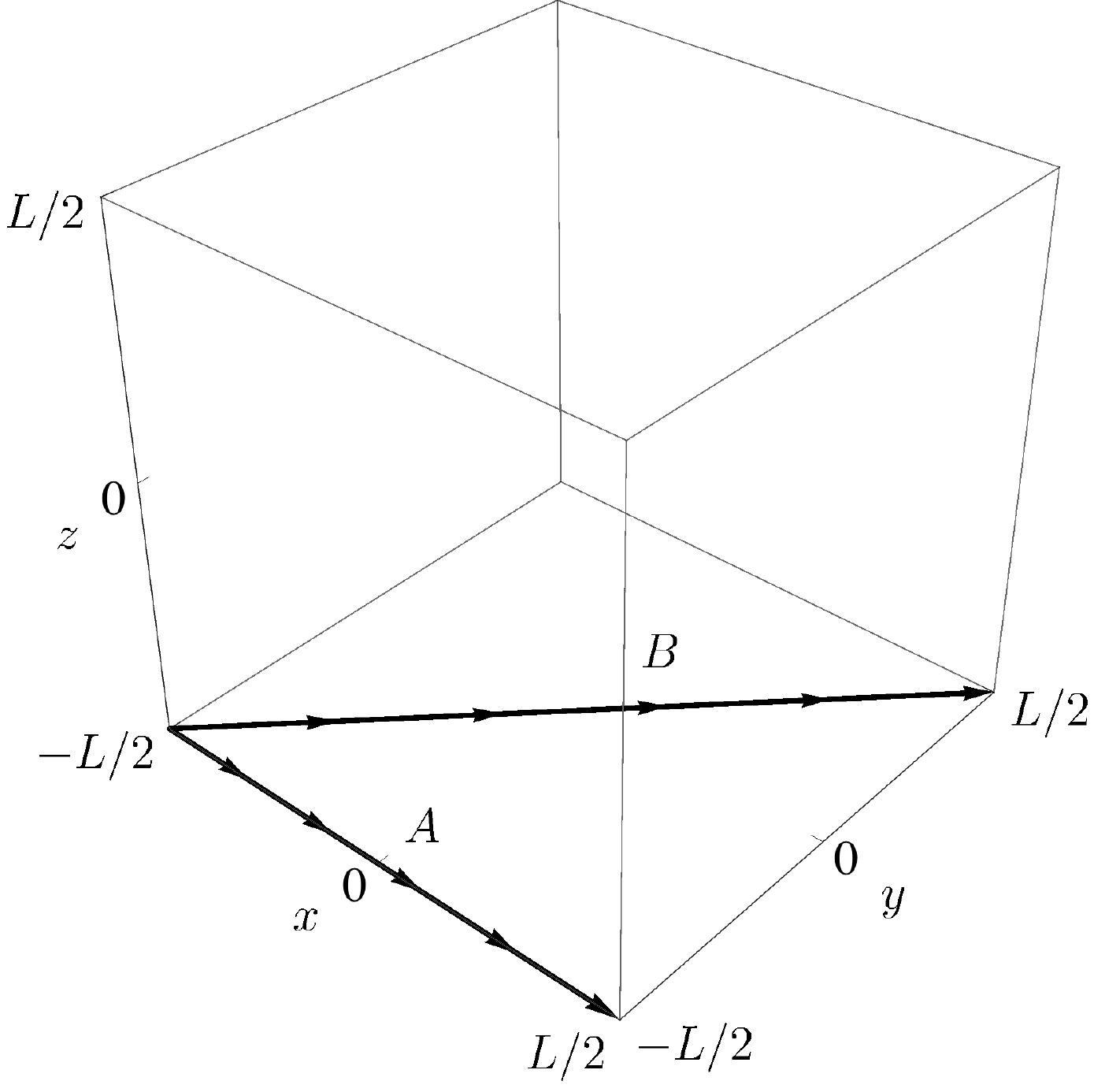} \\
\par\end{centering}
\caption{An illustration of two trajectories, $A$ and $B$, in one of the cubic lattice cells. Both trajectories are geodesic, and run along paths of local rotational symmetry (see Section \ref{sec:arr}). Reproduced from \cite{2017JCAP...03..014B}.\label{bkhg0}}
\end{figure}

The configuration of black hole considered used in this study was the one generated by tiling flat 3-space with an infinite array of cubes. A black hole was then placed at the center of each of these cubes to generate an infinite array that covers the entirety of \Rthree. No exact initial data is known for this type of configuration, so numerical methods were used to solve the constraint equations (as explained in \cite{Yoo:2012jz, 2014CQGra..31c5004B}). The initial data was then evolved numerically, and the optical properties of the two trajectories shown in Fig. \ref{bkhg0} were studied. The first of these (trajectory $A$) follows the edge of one of the cubic lattice cells, while the second (trajectory $B$) follows the diagonal that connects two corners of one of the square cell faces. In both cases these trajectories are contiguous with those of the next neighbouring cell, and so can be extended smoothly whilst remaining far from the masses that reside at the cell centres.

\begin{figure}[t]
\begin{centering}
\includegraphics[width=0.82\textwidth]{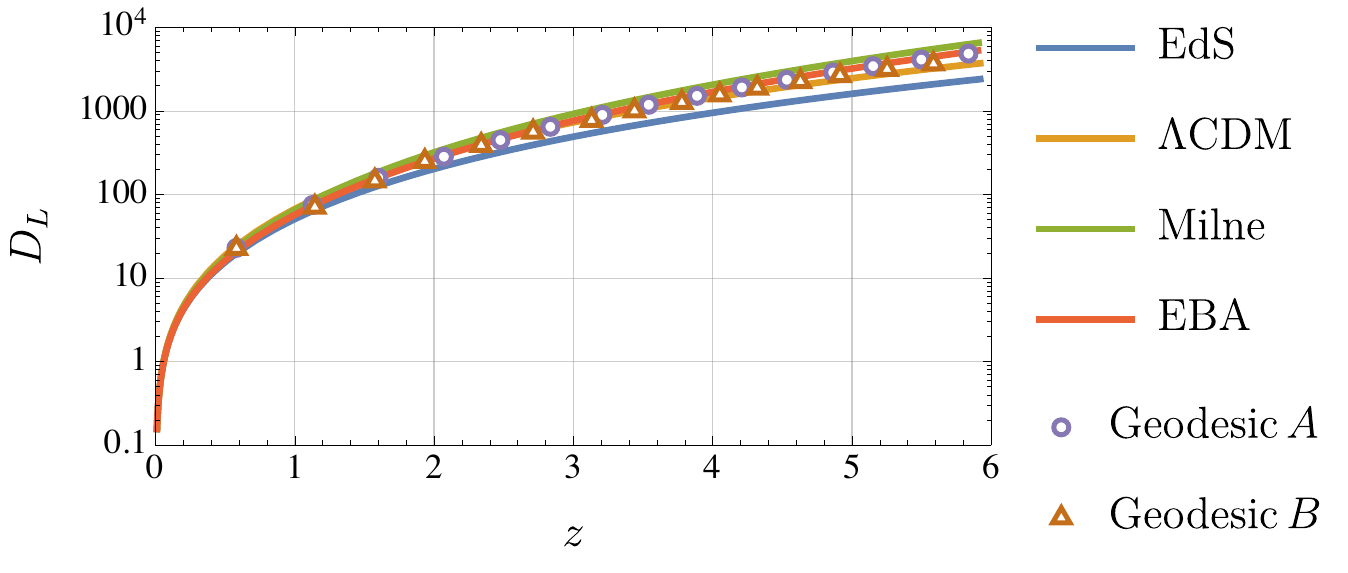} \\
\includegraphics[width=0.57\textwidth]{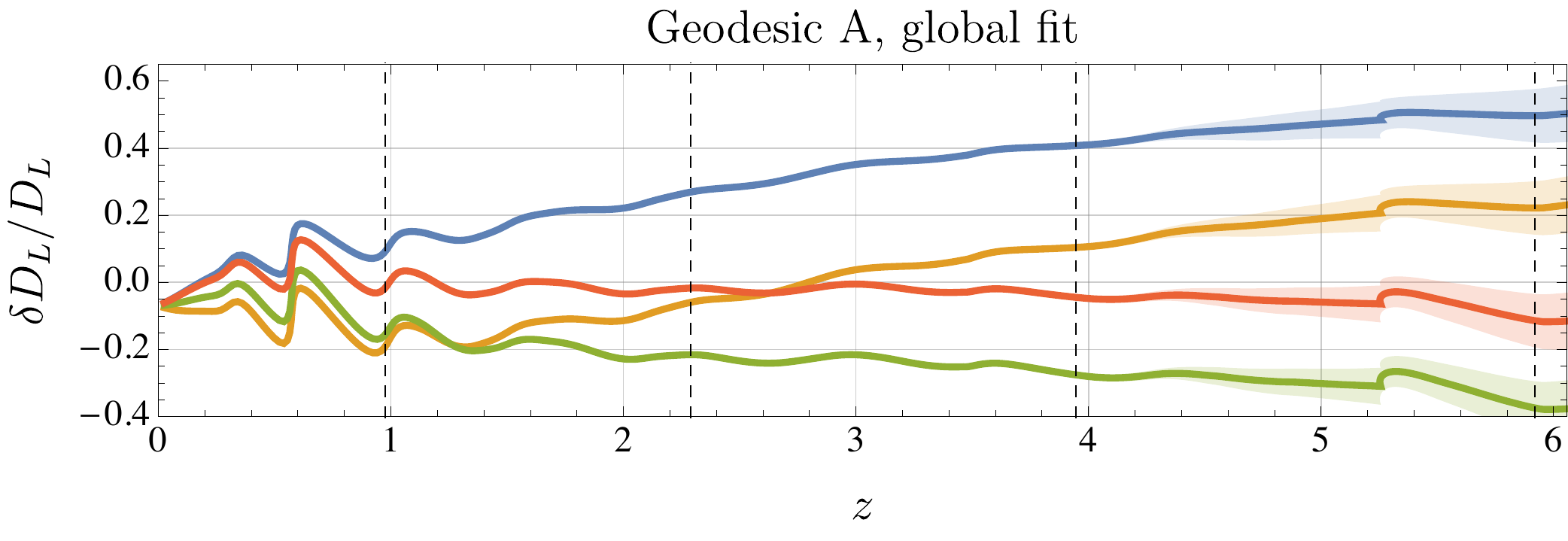} \qquad\qquad\qquad\qquad\\
\includegraphics[width=0.57\textwidth]{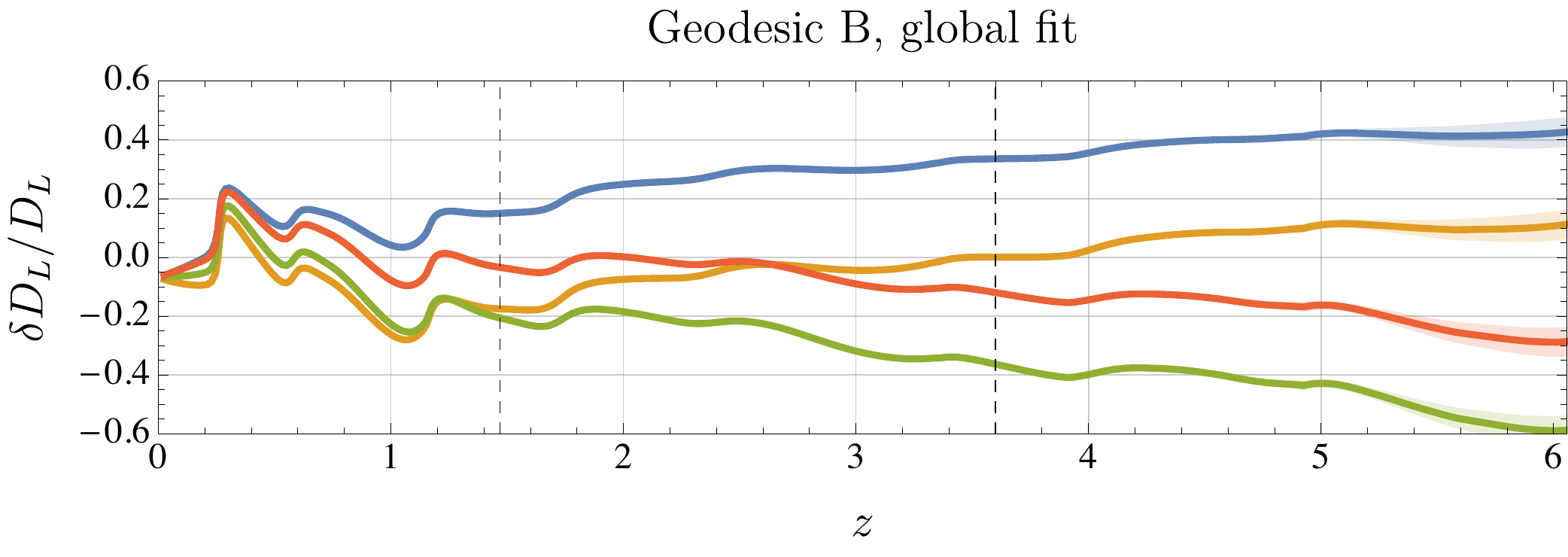} \qquad\qquad\qquad\qquad\\
\par\end{centering}
\caption{Luminosity distance ($D_L$) as a function of redshift ($z$), for observers along trajectories $A$ and $B$ from Figure \ref{bkhg0}. The top panel is a log plot of $D_L(z)$, and also shows the corresponding luminosity distance functions of the Einstein-de Sitter (EdS), $\Lambda$CDM, Milne and Empty Beam Approximation (EBA) models. The relative difference between each of these four additional models, and the optical properties of trajectories $A$ and $B$, are plotted in the second and third panels. In each case the Hubble rate at the present time, $H_0$, was extracted from the rate of change of the length of a cell edge in the lattice spacetime. The $\Lambda$CDM was taken to have $\Omega_{\Lambda}=0.7$ and $\Omega_{m}=0.3$. The error bars are indicated by the shaded regions (when not visible, they are included in the width of the curves). Reproduced from \cite{2017JCAP...03..014B}.\label{bkhg1}}
\end{figure}

The luminosity distance and redshift were then calculated along these two trajectories, using the methods outlined above, with the results shown in Figure \ref{bkhg1}. In both of these figures the redshifts were calculated with respect to a set of observers whose motion was restricted to the lie within the 2-dimensional spaces formed from the evolution of each trajectory, and who were chosen to expand away from each other in a way that is consistent with the large-scale expansion of the spacetime. In a vacuum spacetime of this type, such a set of observers are the closest thing that exists to the frequently-used comoving observers that are routinely used in cosmological models that contain a perfect fluid. Other sets of observers could of course also be introduced, and would result in an effect that could be accounted for by performing a local Lorentz transformation at the point of observation and/or emission.

Together with the results of calculating the luminosity distance along trajectories $A$ and $B$, Figure \ref{bkhg1} also shows the corresponding functions in Einstein-de Sitter, $\Lambda$CDM,  and Milne cosmologies, as well as in the Empty Beam Approximation. The local value of the Hubble rate at $z=0$, in each of these cases, was determined using the rate of expansion of a cell edge in the lattice cosmology. The second and third panels of Figure \ref{bkhg1} show that $D_L(z)$ along trajectory $A$ is well-modeled by the Empty Beam Approximation at all redshifts out to $z=6$ (where the numerical simulation ended). For trajectory $B$, the results of the numerical integration were also well-modeled by the Empty Beam Approximation out to $z \lesssim 3$, but were slightly better approximated by the $\Lambda$CDM model between $3 \lesssim z \lesssim 6$. 

For both trajectories, it can be seen from Figure \ref{bkhg1} that there appear to be significant oscillations in the relative difference of $D_L$ between the lattice models and the comparison models. These oscillations are not numerical errors. Instead, it is believed that they occur from gravitational waves that are introduced into the cosmology by the methods used to prescribe the initial data. For the case of initial data sets that contain asymptotically flat spaces (i.e. those that are usually used for the numerical study of black hole spacetimes), such waves would propagate outwards and rapidly become diffuse enough to be neglected. In the present situation, however, there is nowhere for the waves to radiate away to, as we are dealing with cosmological models that have no accessible asymptotically flat regions. This means that the waves propagate around the cosmology, and cause the oscillatory effects that are visible in Figure \ref{bkhg1}.

The fact that the Empty Beam Approximation accurately reproduces the optical properties along trajectory $A$, and along the low-redshift part of trajectory $B$, is relatively straightforward to understand: It occurs because the spacetime being investigated, and the trajectories in question, display the properties under which the Empty Beam Approximation was formulated. That is, the spacetime is a perfect vacuum, so the Ricci focusing term is entirely absent. As well as this, the local rotational symmetry exhibited around trajectory $A$ means that there is no preferred direction in which shearing of the beam can occur (see Section \ref{sec:arr}). This means trajectory $A$ should exactly satisfy the requirements for the Empty Beam Approximation to be valid (up to the oscillations described above). 

On the other hand, trajectory $B$ can experience shear, due to the smaller number of reflection symmetric surfaces that meet along this curve. This shear accumulates over time, as expected, until at $z \simeq 3$ the empty beam approximation ceases to be the most accurate model for reproducing the results of the numerical integrations. It is interesting that the $\Lambda$CDM model with $\Omega_{\Lambda}=0.7$ gives a decent approximation to trajectory $B$ when $z \gtrsim 3$, although the authors of \cite{2017JCAP...03..014B} also find that larger (or time-dependent) values of $\Omega_{\Lambda}$ provide an even better fit over the full range of redshifts. The possibility of $\Lambda$CDM cosmologies approximating the numerical results displayed above is investigated in more detail in \cite{2017JCAP...03..014B}. 

Finally, the authors of reference \cite{2017JCAP...03..014B} also show that the optical properties of lattice cosmologies do not approach those of FLRW cosmologies, even in the limit where the large-scale expansion does do so. This suggests a possible source of bias, when interpreting observations in a cosmology that is only statistically (and not perfectly) homogeneous and isotropic, which could potentially be of interest for the inference of cosmological parameters.

\subsection{Perturbative spacetime}

Similar analysis to those described above can be performed within the perturbative spacetimes described in Section \ref{sec:approx}. However, in this case the perturbed optical equations are much simpler to solve, and large numbers of photon trajectories can be considered relatively straightforwardly. The perturbative geometry still captures a number of the essential features of the full spacetime, including the lack of Ricci focusing, and the shearing and bending of bundles of null rays as they pass massive objects. In what follows of this section we will describe the results of considering one hundred thousand null geodesics that pass through a cubic lattice constructed in flat space (as was also considered above). This follows the study in \cite{2017JCAP...07..028S}.

\begin{figure}[t!]
\centering
\includegraphics[width=0.5\columnwidth]{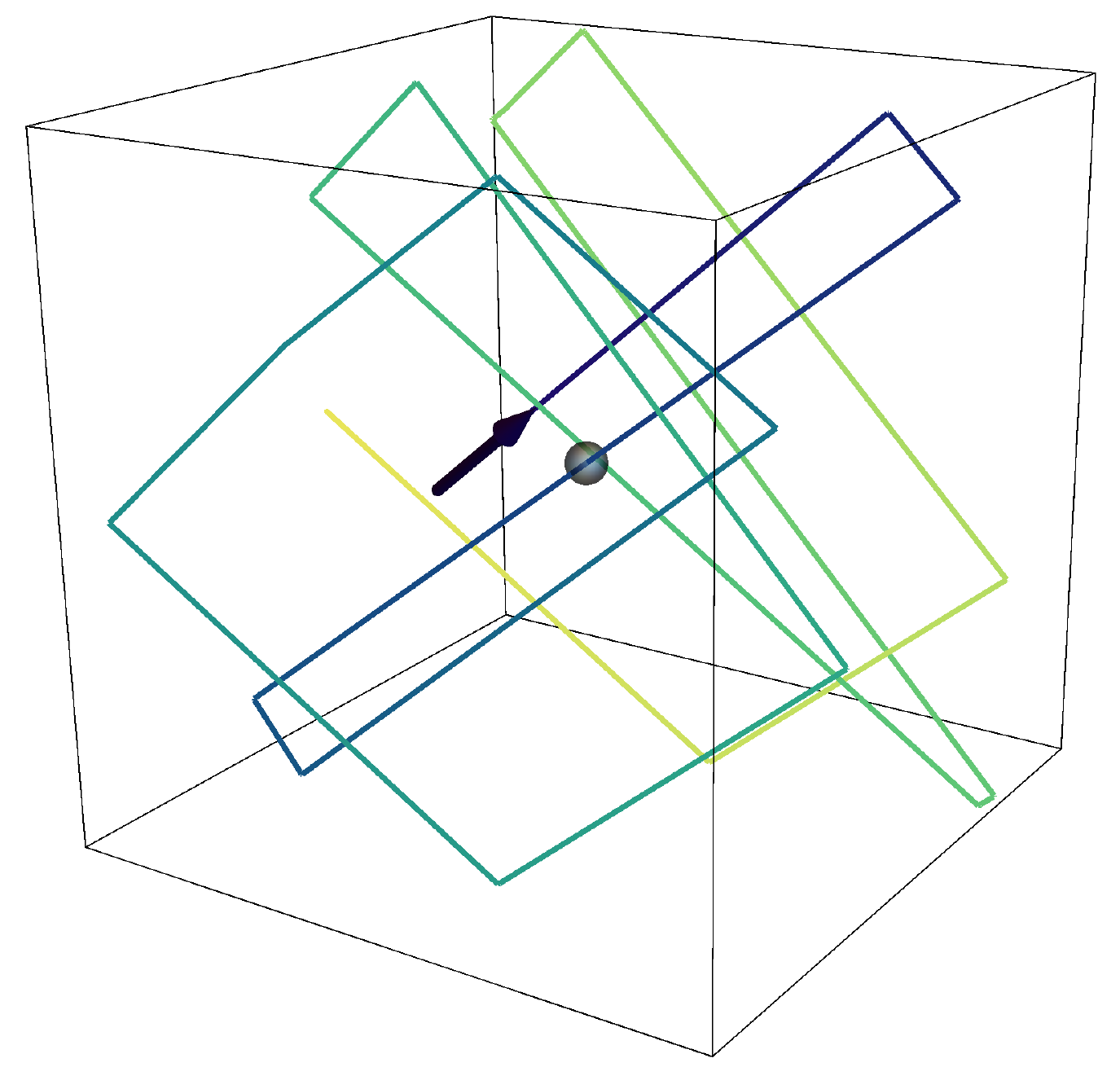}
\caption{A typical realisation of a ray of light, that starts off with an arbitrary position and direction, and that reflects off the walls of the cubic lattice cell. The central sphere indicates a body with radius~$30 \, {\rm kpc}$, in a cell of size~$1 \, {\rm Mpc}$. Reproduced from \cite{2017JCAP...07..028S}.}
\label{fig:ill}
\end{figure}

The method used to collect together this large set of light rays, was to pick a location in the bulk of the lattice cell (away from the walls, the central mass and all lines of local rotational symmetry). Null geodesics were then fired outwards from this location in random directions, and the optical properties of beams of light that follow these geodesics were calculated using perturbed versions of the equations given earlier in this section. When the null geodesic reached a cell face, it was taken to be reflected back into the bulk of the cell and propagated until it hit another cell face. This process was then repeated over and over again until the desired redshift was achieved, as illustrated in Figure \ref{fig:ill}.

The results of this exercise will faithfully reproduce the optical properties of the full lattice universe, as the boundaries between cells is reflection symmetric. This procedure therefore allows one to consider a full cosmology without having to track null geodesics through very large numbers of cells. As in the numerical studies above, the observers were chosen to follow a congruence of curves that expands (in some sense) uniformly, and at the rate given by the large-scale properties of the spacetime. Again, different observers could equally well have been chosen, and the results below could be adapted to such a set by performing local Lorentz transformations at the points of emission and/or observation.

\begin{figure}[t!]
\centering
\includegraphics[width=0.49\columnwidth]{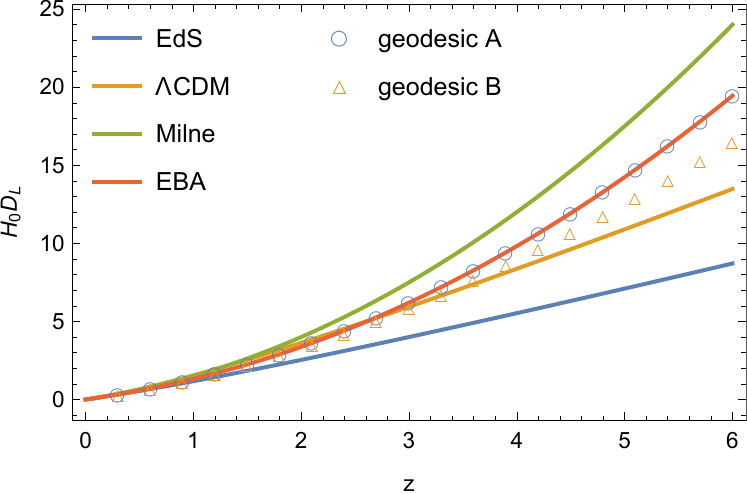}\\
\includegraphics[width=0.5\columnwidth]{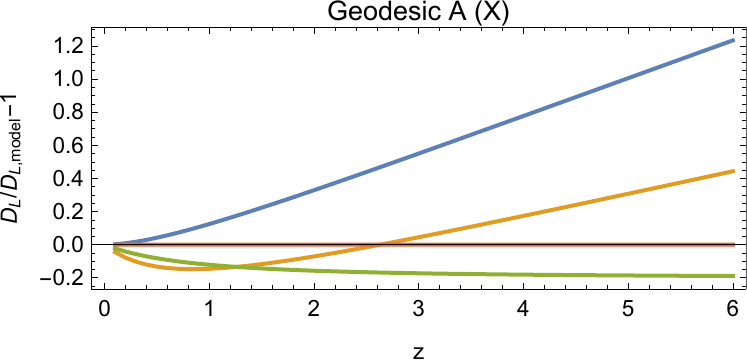}\\
\includegraphics[width=0.5\columnwidth]{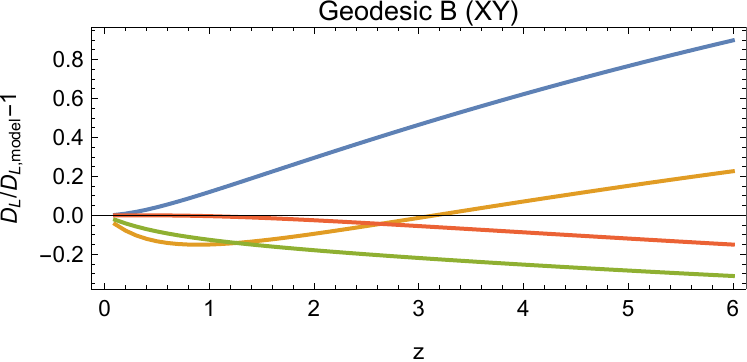}\\
\caption{Luminosity distance ($D_L$) as a function of redshift ($z$), for observers along trajectories $A$ and $B$ from Figure \ref{bkhg0}, in the perturbative spacetime. As in Figure \ref{bkhg1}, the top panel is a plot of $D_L(z)$ along $A$ and $B$, while the second and third panels show the relative difference between $D_L(z)$ along these trajectories or in an Einstein-de Sitter (EdS), $\Lambda$CDM, Milne and Empty Beam Approximation (EBA) model. Again, the $\Lambda$CDM model was taken to have $\Omega_{\Lambda}=0.7$ and $\Omega_{m}=0.3$. The numerical errors in this case are negligibly small. Reproduced from \cite{2017JCAP...07..028S}.}
\label{fig:comp}
\end{figure}

Before considering the statistical properties of the full set of null geodesics described above, it is useful to again consider the two trajectories illustrated in Figure \ref{bkhg0}, this time in the perturbative approximation. The results of this are shown in Figure \ref{fig:comp}, and can be compared to the corresponding results of integrating in the numerical spacetime given in Figure \ref{bkhg1}. The oscillations that were present in the numerical study can now be seen to be absent, as gravitational waves are more easily excluded from the model in this construction. The ability of the Empty Beam Approximation to model the optical properties of trajectory $A$ is much clearer in this case, as is the failure of this approximation to accurately model optics along trajectory $B$.

Having considered single preferred trajectories, let us now consider the results of averaging over large numbers of light rays that shoot out from a single point in random directions. In order to determine such averages accurately, it it necessary to prescribe some details about exactly what it is that lies at the centre of cell. In reference \cite{2017JCAP...07..028S}, two such prescriptions were offered: 

{}\vspace{5pt}
\noindent
(i) {\it Galaxy simulations}: In this case the massive body at the center of the cell is taken to be modeled by the compact bulge of a spiral galaxy. We take such an object to have radius~$R=3\, {\rm kpc}$, and assume it to be perfectly \emph{opaque} (so that any ray that enters into the region $r<3\, {\rm kpc}$ is excluded from the final data set).

{}\vspace{5pt}
\noindent
(ii) {\it Halo simulations}: Here the massive central body is taken to be a dark matter halo with radius~$R=30\, {\rm kpc}$. The halo is assumed to have uniform density, and to be completely \emph{transparent} (so that any light rays that enter it continue to pass through unimpeded).

\begin{figure}[t!]
\centering
\includegraphics[width=0.47\columnwidth]{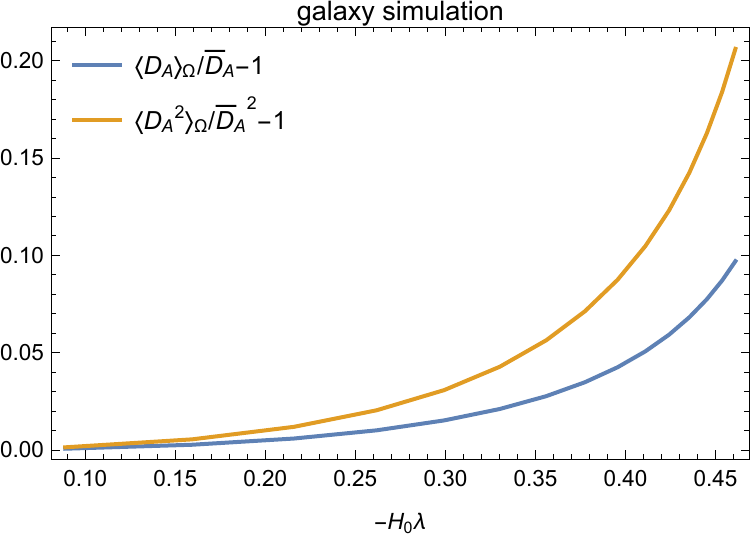}
\hfill
\includegraphics[width=0.5\columnwidth]{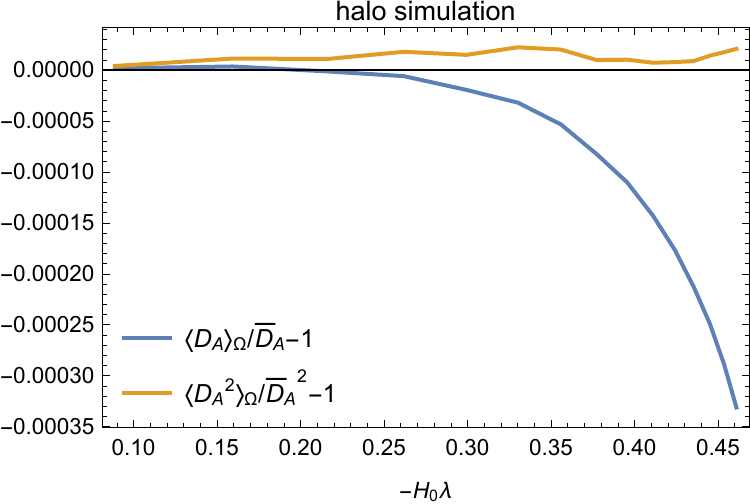}
\caption{The average angular diameter distance, and angular diameter distance squared, for the galaxy simulation (left panel) and the halo simulation (right panel). According to the Weinberg-Kibble-Lieu theorem, it should be found that~$\left\langle D_{A}^2(\lambda) \right\rangle_\Omega/\bar{D}_{A}^2(\lambda) -1 =0$. Reproduced from \cite{2017JCAP...07..028S}.}
\label{fig:DA_DA2}
\end{figure}

\begin{figure}[t!]
\centering
\includegraphics[width=0.47\columnwidth]{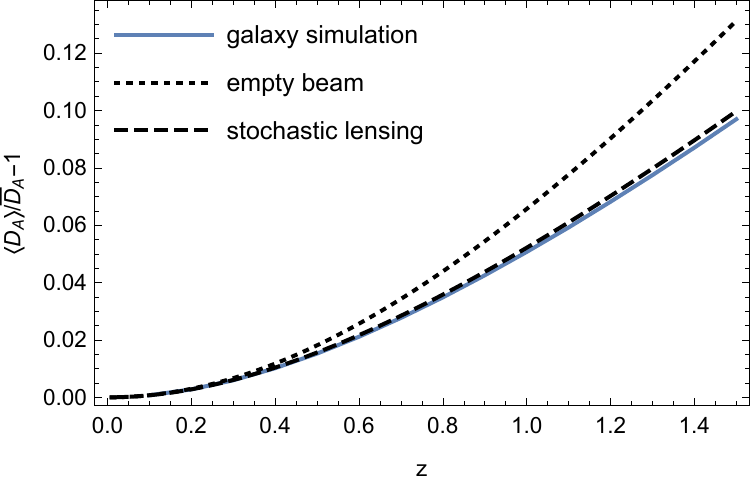}
\hfill
\includegraphics[width=0.5\columnwidth]{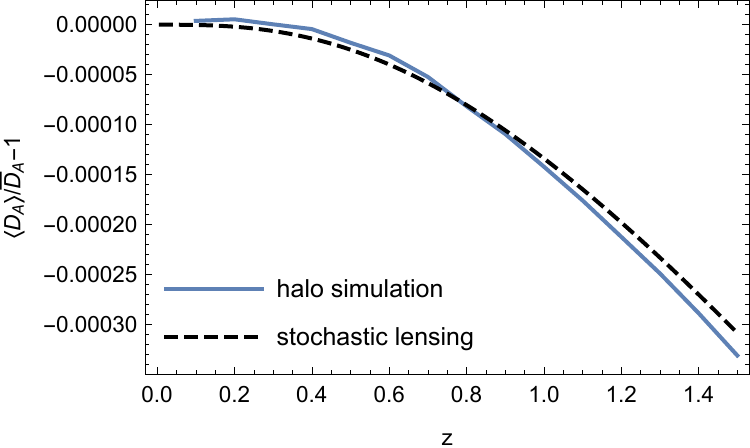}
\caption{Comparison between the sky-averaged angular diameter distance~$\left\langle D_{A} \right\rangle_\Omega$, and the predictions of the Empty Beam Approximation and Stochastic Lensing Formalism, for the galaxy simulations (left panel) and halo simulations (right panel). Reproduced from \cite{2017JCAP...07..028S}.}
\label{fig:bias_DA}
\end{figure}

{}\vspace{5pt}
\noindent
In both of these cases it was found that the effect of the inhomogeneous gravitational field on the average redshift was negligible, when the rays of light were allowed to propagate over large numbers of cells. The average distance measures, however, were found to be significantly different from their FLRW counterparts. The average angular diameter distances are shown for both types of simulation in Figures \ref{fig:DA_DA2} and \ref{fig:bias_DA}. In each of these cases the average is taken over a set of one hundred thousand rays of light, and in each case it is the relative difference from the expected behaviour from FLRW cosmology that is being plotted.

The results shown in Figures \ref{fig:DA_DA2} and \ref{fig:bias_DA} display a number of notable properties. Firstly, the Weinberg-Kibble-Lieu theorem can be shown to be extremely accurate in the halo simulation (right panel of Fig. \ref{fig:DA_DA2}). The small oscillations in the orange line in this plot are due to numerical error, which is estimated at the level of a few parts in $10^5$ in these models. On the other hand, the galaxy simulations (left panel of Fig. \ref{fig:DA_DA2}) do not obey this theorem. This is due to the opaque objects at the centers of the cells, which (despite being very small) are masking the regions of highest curvature. This was found to be enough to bias the inferred estimates of the cosmological parameter $\Omega_{\Lambda}$ by as much as $10\%$, which is quite significant in the modern age of observational cosmology.

Secondly, in Figure \ref{fig:bias_DA}, the average over one hundred thousand light rays is compared to the predictions of the Empty Beam Approximation and the Stochastic Lensing Formalism. The Empty Beam Approximation, while reproducing the findings for the optical properties along the special trajectory $A$, does not appear to be a very good approximation to the averages obtained in either the galaxy or halo simulations. This is because the shear along the vast majority of the trajectories used in these simulations are not free of shear. Instead, the shear accumulates, and focuses the beams of light to an increasing degree as $z$ increases. On the other hand, the Stochastic Lensing Formula does appear to reproduce the results of both simulations to a high degree of accuracy. This lends a lot of credibility to the stochastic lensing approach, but it may be noted that the results of calculating the variance of the distribution around the mean were significantly less successful in the case of galaxy simulations (due to non-Gaussianity in the distribution of the Weyl curvature - see Section 5.2 of \cite{2017JCAP...07..028S} for further details).

Optical properties of approximate, discrete lattices were all studied for the Lindquist-Wheeler model in \cite{Clifton:2009jw,Clifton:2010fr,Clifton:2011mt}, and in the Bruneton-Larena lattice in \cite{Bruneton:2012ru}. The former of these studies showed that the accumulation of shear in the congruence of null geodesics that constitutes a beam of light can be highly non-Gaussian. Rays of light that occasionally venture close to the discrete masses get heavily distorted, and the magnitude of the shear scalar after such a encounter remains high throughout its subsequent evolution. This can lead to magnification of the image, as the shear sources the evolution of the expansion of the bundle, and can even lead to the formation of caustics. Similarly, Bruneton and Larena also found that the expectations from FLRW cosmology can be severely violated if the mass as the centre of each of the cells is allowed to be compact enough, and that this should be expected to lead to a fitting problem.

\section{Conclusions}

Black-hole lattices are valuable toy models, in which the effect of strong-field, non-linear 
inhomogeneities on a cosmological spacetime can be investigated in exact, fully relativistic terms.
Though the first approach to the study of BHLs was proposed over sixty years ago,
the last decade has witnessed a tremendous burst of developments in their modelling and understanding:
this has involved generalised analytic constructions, along with exact initial data and their numerical evolution, as well as ray tracing across the resulting spacetimes. The results of this effort have shed considerable light on many of the issues involving the inclusion of non-linear structures in cosmology, 
and have enabled the calculation of cosmologically relevant effects, such as the kinematical
backreaction, the dressing of mass and other physical parameters, and the biases in
optical measurements that inhomogeneous gravitational fields can produce. This review
presents the current state of the art in each of these subjects.

In particular, we have described the conditions under which exact multi-black-hole 
solutions of Einstein's equations can be 
constructed (Section~\ref{sec:arr}). There are now a large variety of known instances of such
BHLs, involving different global topologies, stress-energy content, and multi-scale 
configurations. We have then reviewed how the techniques of Numerical Relativity can be used
to evolve these arrangements in time, and to develop the initial data into full spacetimes
(Section~\ref{sec:evol}). This provides a unique tool to model the kinematical behaviour
of these systems, and observe how their properties change as they expand or collapse.
An essential aspect of these models is whether, and under which conditions, lattices
of smaller and smaller mass-to-spacing ratio behave as cosmologies filled with smooth
fluids. This continuum limit is explored in Section~\ref{sec:cont}. 
In Section~\ref{sec:effect}, we then consider the extent to which the dynamical properties of 
a BHL differ from the bare superposition of its constituents, an effect we dub
\emph{dressing} (to emphasise the analogy with the coarse-graining of other physical
theories).
Finally, we discuss how these spacetimes appear to observers immersed in them, 
comparing observables such as the luminosity-distance-to-redshift relationship 
in BHLs and FLRW cosmologies. This is done in Section~\ref{sec:opt}.

Each of the above sections presents a different perspective into the effects of 
strong inhomogeneity on the large-scale behaviour of spacetime. However, a few leitmotifs can be easily identified: for instance, it is clear
that the conditions for the existence of spatial hypersurfaces with multiple black
holes in a periodic arrangement are closely related to the conditions for the 
existence of spatially homogeneous and isotropic hypersurfaces in a FLRW cosmology,
a fact that directly influences the BHL constructions of Section~\ref{sec:arr}
(and has potential ramifications on other aspects of these models). 
Furthermore, it is apparent that the mapping between the BHL class of cosmologies and the FLRW models
can be very different depending on the quantity used to construct it, an expression
of the fitting problem that occurs generically in General Relativity~\cite{Ellis:1987zz}. Overall,
BHLs have proven a powerful tool to not only create inhomogeneous cosmological
models, but also to gain insight into the way gravitational structures 
assemble, and their cosmological properties emerge, from the collective interaction of 
many individual bodies in General Relativity.

\section*{Acknowledgements}
TC and JD are supported by the STFC. MK was supported by the National Science Centre, Poland (NCN) via the SONATA BIS programme, grant No~2016/22/E/ST9/00578 for the project
\emph{``Local relativistic perturbative framework in hydrodynamics and general relativity and its application to cosmology''}.

\bibliographystyle{iopart-num}
\bibliography{refs}

\end{document}